\documentclass[11pt,a4paper]{amsart}
\usepackage{amssymb}
\usepackage[dvips,bookmarks,colorlinks=true,linkcolor=blue,urlcolor=red]{hyperref}
\numberwithin{equation}{section}
\newcommand{\D}{\displaystyle}

\newcommand{\car}{\mathbf{1}}
\newcommand{\R}{{\mathbb R}}

\newcommand{\Z}{{\mathbb Z}}
\newcommand{\N}{{\mathbb N}}
\newcommand{\C}{{\mathbb C}}

\newcommand{\E}{{\mathbb E}}

\renewcommand{\P}{{\mathbb P}}
\newcommand{\esp}{{\mathbb E}}
\newcommand{\tr}{{\rm tr}\,}

\newcommand{\vers}{\operatornamewithlimits{\to}}
\theoremstyle{plain}
\newtheorem{Th}{Theorem}[section]
\newtheorem{Le}{Lemma}[section]
\newtheorem{Pro}{Proposition}[section]

\theoremstyle{definition}
\newtheorem{Rem}{Remark}[section]
\title[Inverse tunneling and spectral statistics]{Inverse tunneling
  estimates and applications to the study of spectral statistics of
  random operators on the real line}
\author{Fr{\'e}d{\'e}ric Klopp} 
\address[Fr{\'e}d{\'e}ric Klopp]{IMJ, UMR CNRS 7586
 Universit{\'e} Pierre et Marie Curie
 Case 186, 4 place Jussieu
 F-75252 Paris cedex 05
 France}
\email{\href{mailto:klopp@math.jussieu.fr}{klopp@math.jussieu.fr}}
\begin{document}
%
\begin{abstract}
  We present a proof of a Minami type estimate for one dimensional
  random Schr{\"o}dinger operators valid at all energies in the
  localization regime provided a Wegner estimate is known to hold. The
  Minami type estimate is then applied to two models to obtain
  results on their spectral statistics.\\
  The heuristics underlying our proof of Minami type estimates is that
  close by eigenvalues of a one-dimensional Schr{\"o}dinger operator
  correspond either to eigenfunctions that live far away from each
  other in space or they come from some tunneling phenomena. In the
  second case, one can undo the tunneling and thus construct
  quasi-modes that live far away from each other in space.
  \vskip.5cm\noindent \textsc{R{\'e}sum{\'e}.} Nous d{\'e}montrons une in{\'e}galit{\'e}
  de type Minami pour des op{\'e}rateurs de Schr{\"o}dinger al{\'e}atoires
  uni-dimensionnel dans toute la r{\'e}gion localis{\'e}e si une estim{\'e}e de
  Wegner est connue. Cette estim{\'e}e de type de Minami est alors
  appliqu{\'e}e pour obtenir les statistiques spectrales pour deux
  mod{\`e}les.\\
  L'heuristique qui guide ce travail est que des valeurs propres
  proches pour un op{\'e}rateur de Schr{\"o}dinger sur un intervalle sont
  soient localis{\'e}es loin l'une de l'autre soit sont la cons{\'e}quence
  d'un ph{\'e}nom{\`e}ne d'``effet tunnel''. Dans le second cas, on peut, en
  ``d{\'e}faisant'' cet effet tunnel construire des quasi-modes qui sont
  localis{\'e}s loin l'un de l'autre.
\end{abstract}
\thanks{The author is partially supported by the grant
  ANR-08-BLAN-0261-01.}
\setcounter{section}{-1}
\maketitle
\section{Introduction}
\label{intro}
Consider the following two random operators on the real line
\begin{itemize}
\item the Anderson model
  \begin{equation}
    \label{eq:60}
    H^A_\omega=-\frac{d^2}{dx^2}+W(\cdot)+\sum_{n\in\Z}\omega_n V(\cdot-n)
  \end{equation}
  where
  \begin{itemize}
  \item $W:\ \R\to\R$ is a bounded, continuous, $\Z$-periodic
    function;
  \item $V:\ \R\to\R$ is a bounded, continuous, compactly supported,
    non negative, not identically vanishing function;
  \item $(\omega_n)_{n\in\Z}$ are bounded i.i.d random variables, the
    common distribution of which admits a continuous density.
  \end{itemize}
\item the random displacement model
  \begin{equation}
    \label{eq:61}
    H^D_\omega=-\frac{d^2}{dx^2}+\sum_{n\in\Z} V(\cdot-n-\omega_n)
  \end{equation}
  where
  \begin{itemize}
  \item $V:\ \R\to\R$ is a smooth, even function that has a fixed sign
    and is compactly supported in $(-r_0,r_0)$ for some $0<r_0<1/2$;
  \item $(\omega_n)_{n\in\Z}$ are bounded i.i.d random variables, the
    common distribution of which admits a density supported in
    $[-r,r]\subset [-1/2+r_0,1/2-r_0]$, that is continuously
    differentiable in $[-r,r]$ and which support contains $\{-r,r\}$.
  \end{itemize}
\end{itemize}
Let $\bullet\in\{A,D\}$. For $L>0$, consider $H^\bullet_{\omega,L}$
the operator $H^\bullet_\omega$ restricted to the interval
$[-L/2,L/2]$ with Dirichlet boundary conditions. The spectrum of this
operator is discrete and accumulates at $+\infty$ ; we denote it by
\begin{equation*}
  E^\bullet_1(\omega,L)< E^\bullet_2(\omega,L)\leq \cdots\leq
  E^\bullet_n(\omega,L)\leq \cdots
\end{equation*}
It is well known (see e.g.~\cite{MR94h:47068}) that, $\omega$ almost
surely, the limit
\begin{equation}
  \label{eq:19}
  N^\bullet(E)=\lim_{L\to+\infty}\frac{\#\{n;\
    E^\bullet_n(\omega,L)\leq E\}}L
\end{equation}
exists and is independent of $\omega$. $N^\bullet$ is the {\it
  integrated density of states } of the operator
$H^\bullet_\omega$. This non decreasing function is the distribution
function of a non negative measure, say, $dN^\bullet$ supported on
$\Sigma^\bullet$, the almost sure spectrum of $H^\bullet_\omega$.\\
Moreover, it is known that, under our assumptions, $N^A$ is Lipschitz
continuous on $\R$ (see~\cite{MR2362242}) and there exists $\tilde
E^D\in(\inf\Sigma^D,+\infty)$ such that, for any $\alpha\in(0,1)$,
$N^D$ is Lipschitz continuous on $(-\infty,\tilde E^D)$ (see
Theorem~\ref{thm:wegner} in section~\ref{sec:local-model-h_d}).\\
For a fixed energy $E_0$, one defines the {\it locally unfolded
  levels} to be the points
\begin{equation*}
  \xi^\bullet_n(E_0,\omega,L)=L\,
  [N^\bullet(E^\bullet_n(\omega,L))-N^\bullet(E_0)].
\end{equation*}
Out of these points form the point process
\begin{equation*}
  \Xi^\bullet(\xi; E_0,\omega,L) = 
  \sum_{n\geq1} \delta_{\xi^\bullet_n(E_0,\omega,L)}(\xi),
\end{equation*}
The local level statistics are described by
\begin{Th}
  \label{thr:4}
  There exists an energy $\inf\Sigma^\bullet<E^\bullet\leq\tilde
  E^\bullet$ and such that, if $E_0\in
  (-\infty,E^\bullet)\cap\Sigma^\bullet$ satisfies, for some
  $\rho\in[1,4/3)$, one has
  \begin{equation}
    \label{eq:58}
    \forall a>b,\ \exists C>0,\ \exists \varepsilon_0>0,
    \forall\varepsilon\in(0,\varepsilon_0),\
    |N^\bullet(E_0+a\varepsilon)-N^\bullet(E_0+b\varepsilon)|\geq C
    \varepsilon^{\rho}
  \end{equation}
  then, when $L\to+\infty$, the point process $\Xi(E_0,\omega,L)$
  converges weakly to a Poisson process on $\R$ with intensity the
  Lebesgue measure.
\end{Th}
\noindent One easily checks that, if $E_0$ is such that $E\mapsto
N(E)$ is differentiable at $E_0$ and its derivative is positive,
then~\eqref{eq:58} is satisfied. For both models, this is the case for
Lebesgue almost all points in
$[\inf\Sigma^\bullet,E^\bullet)\cap\Sigma^\bullet$. To the best of our
knowledge, Theorem~\ref{thr:4} gives the first instance of a model
that is not of alloy type for which local Poisson statistics have been
proved.\vskip.2cm
\noindent As is to be expected from
e.g.~\cite{MR673165,MR97d:82046,Ge-Kl:10} and as we shall see in
section~\ref{sec:setting-results}, the local Poisson statistics
property holds over the localized region of the spectrum i.e. the
energy $E^\bullet$ is the energy such that $H_\omega^\bullet$ is
localized in $(-\infty,E^\bullet)$. In particular, the conclusions of
Theorem~\ref{thr:4} also holds in any region of localization of
$H_\omega^\bullet$ where a Wegner type estimate is known to hold.\\
When $\bullet=A$,, in section~\ref{sec:local-model-h_om}, extending
the analysis done in~\cite{0912.3568}, we show that the localized
region (in the sense of (Loc)) extends over the whole real axis. Thus,
we can take $E^A=+\infty$.\\
When $\bullet=D$, it was proved in~\cite{MR1804511} that the
localization region also extends over the whole real axis except for
possibly a discrete set; here localization did not mean (Loc) but a
weaker statement, namely, that the spectrum is pure point associated
to exponentially decaying eigenvalues. The analysis
in~\cite{MR1804511} works under assumptions less restrictive than
those made above. In section~\ref{sec:local-model-h_d}, extending the
analysis done in~\cite{1007.2483}, we show that $H_\omega^D$ satisfies
(Loc) in some neighborhood of $\inf\Sigma^D$.\\
Note that, to obtain the local Poisson statistics near an energy
$E_0$, we do not require the density of states, i.e. the derivative of
$N^\bullet$, not to vanish at $E_0$; we only require that $N^\bullet$
not be too flat near $E_0$.\\
Following the ideas of~\cite{Ge-Kl:10,Kl:10a}, using the Minami type
estimates that we present in section~\ref{sec:setting-results}, we can
obtain a host of other results on the asymptotics of the statistics of
the eigenvalues of the random operator $H^\bullet_{\omega,L}$ in the
localized regime. We now give a few of those.\\
Fix $\bullet\in\{A,D\}$. For $J=[a,b]$, a compact interval s.t.
$|N^\bullet(J)|:=N^\bullet(b)-N^\bullet(a)>0$ and a fixed
configuration $\omega$, consider the point process
\begin{equation*}
  \Xi^\bullet_J(\omega,t,L) = 
  \sum_{E^\bullet_n(\omega,L)\in J}
  \delta_{|N^\bullet(J)|L[N^\bullet_J(E^\bullet_n(\omega,L))-t]}
\end{equation*}
under the uniform distribution in $[0,1]$ in $t$; here we have set
\begin{equation*}
  N^\bullet_J(\cdot):=
  \frac{N^\bullet(\cdot)-N^\bullet(a)}{N^\bullet(b)-N^\bullet(a)}.   
\end{equation*}
This process was introduced in~\cite{MR2352280,Mi:11}; we refer to
these papers for more references, in particular, for references to the
physics literature. The values
$(N_J^\bullet(E^\bullet_n(\omega,L)))_{n\geq1}$ are called the {\it
  $J$-unfolded eigenvalues} of the operator $H^\bullet_{\omega,L}$.\\
Following~\cite{Kl:10a}, one proves
\begin{Th}
  \label{thr:9}
  Fix $J=[a,b]\subset(-\infty,E^\bullet)\cap\Sigma^\bullet$ a compact
  interval such that $|N^\bullet(J)|>0$. Then, $\omega$-almost surely,
  as $L\to+\infty$, the probability law of the point process
  $\Xi^\bullet_J(\omega,\cdot,L)$ under the uniform distribution
  $\car_{[0,1]}(t)dt$ converges to the law of the Poisson point
  process on the real line with intensity $1$.
\end{Th}
\noindent As is shown in~\cite{Mi:11}, Theorem~\ref{thr:9} implies the
convergence of the unfolded level spacings distributions for the
levels in $J$.  More precisely, define the $n$-th unfolded eigenvalue
spacings
\begin{equation}
  \label{eq:64}
  \delta N^\bullet_n(\omega,L)=L|N^\bullet(J)|
  (N_J^\bullet(E_{n+1}(\omega,L))-
  N^\bullet(E_n(\omega,L)))\geq0.
\end{equation}
Define the empirical distribution of these spacings to be the random
numbers, for $x\geq0$
\begin{equation}
  \label{eq:66}
  DLS^\bullet(x;J,\omega,L)=\frac{\#\{j;\ E^\bullet_n(\omega,L)\in
    J,\ \delta N^\bullet_n(\omega,L)\geq x\}
  }{N^\bullet(J,\omega,L)}
\end{equation}
where $N^\bullet(J,\omega,L):=\#\{E^\bullet_n(\omega,L)\in J\}=
|N^\bullet(J)|L(1+o(1))$ as $L\to+\infty$ (see e.g.~\cite{Ge-Kl:10a}).
\begin{Th}
  \label{thr:10}
  Under the assumptions of Theorem~\ref{thr:9}, $\omega$-almost
  surely, as $L\to+\infty$, $DLS^\bullet(x;J,\omega,L)$ converges
  uniformly to the distribution $x\mapsto e^{-x}$.
\end{Th}
\noindent One can also obtain results for the eigenvalues themselves
i.e. when they are not unfolded; we refer to~\cite{Ge-Kl:10,Kl:10a}
for more details.\\
Finally we turn to results on level spacings that are local in energy
(in the sense of Theorem~\ref{thr:4}). Fix
$E_0\in(-\infty,E^\bullet)\cap\Sigma^\bullet$. Pick $I_L=[a_L,b_L]$, a
small interval centered near $0$. With the same notations as above
(see~\eqref{eq:64}), define the empirical distribution of these
spacings to be the random numbers, for $x\geq0$
\begin{equation}
  \label{eq:20}
  DLS^\bullet(x;I_L,\omega,L)=\frac{\#\{j;\ E^\bullet_j(\omega,L)-E_0\in
    I_L,\ \delta N^\bullet_j(\omega,L)\geq x\}
  }{N^\bullet(E_0+I_L,L,\omega)}.
\end{equation}
We prove
\begin{Th}
  \label{thr:5}
  Assume that $E_0\in[\inf\Sigma^\bullet,E^\bullet)$.  Fix $(I_L)_L$ a
  decreasing sequence of intervals such that
  $\displaystyle\sup_{I_L}|x|\vers_{L\to+\infty}0$.  Assume that, for
  some $\delta>0$ and $\tilde\rho\in[1,4/3)$, one has
  \begin{gather}
    \label{eq:59}
    N(E_0+I_L)\cdot|I_L|^{-\tilde\rho}\geq1,\\
    \intertext{ and }
    \label{eq:18}
    \quad L^{1-\delta}\cdot N(E_0+I_L) \vers_{L\to+\infty}
    +\infty,\quad\frac{N(E_0+I_{L_{L+o(L)}})}{N(E_0+I_{L_L})}
    \vers_{L\to+\infty}1.
  \end{gather}
  Then, with probability $1$, as $L\to+\infty$, $DLS(x;I_L,\omega,L)$
  converges uniformly to the distribution $x\mapsto e^{-x}$, that is,
  with probability $1$,
  \begin{equation}
    \label{eq:67}
    \sup_{x\geq0}\left|DLS(x;I_L,\omega,L)
      -e^{-x}\right|\vers_{L\to+\infty}0.
  \end{equation}
\end{Th}
\noindent As condition~\eqref{eq:58}, condition~\eqref{eq:59} is
satisfied for $\tilde\rho=1$ for almost every
$E_0\in[\inf\Sigma^\bullet,E^\bullet)$.  Condition~\eqref{eq:18} on
the intervals $(I_L)_L$ ensures that they contain sufficiently many
eigenvalues for the empirical distribution to make sense and that this
number does not vary too wildly when one slightly changes the size of
$I_L$.\\
The main technical result of the present paper that we turn to below
entail a number of other consequences about the spectral statistics in
the localized region. We refer to~\cite{Ge-Kl:10,Kl:10a} for more such
examples and more references.
\section{The setting and the results}
\label{sec:setting-results}
Let us now turn to the main result of this paper. It concerns random
operators on the real line and consist in Minami type estimates valid
for all energies in the localization region of general one dimensional
random operators satisfying a Wegner estimate. It can be summarized as
follows:
\begin{itemize}
\item for one dimensional random Schr{\"o}dinger operators, in the
  localization region, a Wegner estimate implies a Minami estimate.
\end{itemize}
The statement does not depend on the specific form of the random
potential.\\
Let us start with a description of our setting. From now on, on
$L^2(\R)$, we consider random Schr{\"o}dinger operators of the form
\begin{equation}
  \label{eq:7}
  H_\omega u=-\frac{d^2}{dx^2}u+q_\omega\,u
\end{equation}
where $q_\omega$ is an almost surely bounded $\Z^d$-ergodic random
potential.
\begin{Rem}
  \label{rem:4}
  The boundedness assumption may be relaxed so as to allow local
  singularities and growth at infinity. We make it to keep our proofs
  as simple as possible.
\end{Rem}
\noindent It is well known (see e.g.~\cite{MR94h:47068}) that
$H_\omega$ then admits an integrated density of states, say, $N$, and,
an almost sure spectrum, say, $\Sigma$. We now fix $I$ an open
interval in $\Sigma$ and the subsequent assumptions and statements
will be made on energies in $I$.\\
Let $H_{\omega}(\Lambda)$ be the random Hamiltonian $H_\omega$
restricted to the interval $\Lambda:=[0,L]$ with periodic boundary
conditions.\\
We now state our main assumptions and comment on the validity of these
assumptions for the models $H_\omega^{A,D}$ defined respectively
in~\eqref{eq:60} and~\eqref{eq:61}.\\
Our first assumption will be a independence assumption for local
Hamiltonians that are far away from each other, that is,
\begin{description}
\item[(IAD)] There exists $R_0>0$ such that for
  dist$(\Lambda,\Lambda')> R_0$, the random Hamiltonians
  $H_\omega(\Lambda)$ and $H_\omega(\Lambda')$ are independent.
\end{description}
\begin{Rem}
  \label{rem:3}
  This assumption may be relaxed to asking some control on the
  correlations between the random Hamiltonians restricted to different
  cubes. To keep the proofs as simple as possible, we assume (IAD).
\end{Rem}
Next, we assume that
\begin{description}
\item[(W)] a Wegner type estimate holds i.e. there exists $C>0$,
  $s\in(0,1]$ and $\rho\geq1$ such that, for $J\subset I$, and
  $\Lambda$, an interval in $\R$, one has
  \begin{equation}
    \label{eq:1}
    \E\left[\text{tr}(\car_J(H_\omega(\Lambda)))
    \right]\leq C|J|^s|\Lambda|^\rho.
  \end{equation}
  Here, $|\cdot|$ denotes the length of the interval $\cdot$.
\end{description}
\begin{Rem}
  \label{rem:2}
  In many cases e.g. for the operators $H_\omega^{A,D}$, assumption
  $(W)$ is known to hold for $s=1$ and $\rho=1$. In the case of
  $H_\omega^{A}$, we can take $I=\Sigma^A$ (see e.g.~\cite{MR2362242}).\\
  For Anderson type Hamiltonians with single site potentials that are
  not of fixed sign, Wegner estimates with arbitrary $s\in(0,1)$ and
  $\rho=1$ have been proved near the bottom of the spectrum and at
  spectral edges (see~\cite{MR95m:82080,MR1934351}).\\
  In the case of $H_\omega^{D}$, it holds for any $s\in(0,1)$ and
  $\rho=1$ near the infimum of $\Sigma^D$ (see
  section~\ref{sec:local-model-h_d} and~\cite{1007.2483}).
\end{Rem}
The second assumption crucial to our study is the existence of a
localization region to which $I$ belongs i.e. we assume
\begin{description}
\item[(Loc)] for any $\xi\in(0,1)$, one has
  \begin{equation}
    \label{eq:84}
    \sup_{\substack{L>0\\\text{supp} f\subset I \\ |f|\leq1}}\,
    \esp\left(\sum_{n\in\Z} \ e^{|n|^\xi}\,
      \|\car_{[-1/2,1/2]}f(H_\omega(\Lambda_L))
      \car_{[n-1/2,n+1/2]}\|_2\right)<+\infty.
  \end{equation}
  Here, the supremum is taken over all Borel functions $f:\R\to \C$
  which satisfy $|g|\le 1$ pointwise.
\end{description}
\begin{Rem}
  \label{rem:1}
  For the models $H_\omega^{A,D}$, the spectral theory has been
  studied under various assumptions on $V$ and
  $(\omega_\gamma)_\gamma$ (see
  e.g.~\cite{MR1102675,MR1915036,MR1785430,MR2587048}). The existence
  of a region of localized states is well known and, in many cases,
  this region extends over the whole spectrum. In the case of
  $H_\omega^A$, in~\cite{0912.3568}, this is proved under a more
  restrictive support condition on $V$, namely, that the support of
  $V$ is contained in $(-1/2,1/2)$; that this condition can be removed
  is proved in section~\ref{sec:local-model-h_om}. Actually, for this
  model we get a stronger form of (Loc), namely, for any $I\subset\R$
  compact, there exists $\xi=\xi_I>0$ such that
  \begin{equation}
    \label{eq:81}
    \sup_{\substack{L>0\\\text{supp} f\subset I \\ |f|\leq1}}\,
    \esp\left(\sum_{n\in\Z} \ e^{\xi|n|}\,
      \|\car_{[-1/2,1/2]}f(H_\omega(\Lambda_L))
      \car_{[n-1/2,n+1/2]}\|_2\right)<+\infty.
  \end{equation}
  In the case of model $H_\omega^{D}$, localization (in a sense weaker
  than assumption (Loc) above) has been proved at all energies except
  possibly at a discrete set (see~\cite{MR1804511}). In dimension
  $d\geq2$, localization at the bottom of the spectrum for
  $H_\omega^{D}$ has been proved in~\cite{1007.2483}. This proof does
  not work directly in dimension $d=1$. In
  section~\ref{sec:local-model-h_d}, we show prove that, under our
  assumptions, $H_\omega^{D}$ satisfies (Loc) at the bottom of the
  spectrum. \\
  There are other random models for which localization (in a possibly
  weaker sense than (Loc) above) has been proved e.g. the Russian
  model (\cite{MR0470515}), the Bernoulli Anderson model
  (\cite{MR1915036,MR2180453,2011arxiv1105.0213G}), the Poisson model
  (\cite{MR1363537,MR2314108}), more general displacement models
  (\cite{MR1804511}), matrix valued models (\cite{MR2525594}),
  etc. For many of these models, the validity of (W) is still an open
  question.
\end{Rem}
\subsection{A Minami type estimate in the localization region}
\label{sec:impr-minami-estim}
Our main technical result is the following Minami type estimate
\begin{Th}
  \label{thr:1}
  Assume (W) and (Loc). Fix $J$ compact in $I$ the region of
  localization. Then, for $\eta>1$, $\beta>\max(1+4s,\rho)$ and
  $\rho'>\rho$ (recall that $\rho$ and $s$ are defined in (W)), there
  exists $L_{\eta,\beta,\rho'}>0$ and $C=C_{\eta,\beta,\rho'}>0$ s.t.,
  for $E\in J$, $L\geq L_{\eta,\beta,\rho'}$ and
  $\varepsilon\in(0,1)$, one has
  \begin{multline}
    \label{eq:8}
    \sum_{k\geq2}\P\left(\tr[\car_{[E-\varepsilon,E+\varepsilon]}
      (H_\omega(\Lambda_L))]\geq k\right)\\\leq C\left[\left(
        \varepsilon^s\,L\,\ell^\beta+e^{-\ell/8}\right)^2
      e^{C\,\varepsilon^s\, L\,\ell^{\rho'}} +e^{-s\ell/9}\right].
  \end{multline}
  where $\ell:=(\log L)^\eta$.
\end{Th}
\noindent The estimate~\eqref{eq:8} only becomes useful when
$\varepsilon^sL$ is small; as $\rho\geq1$, this is also the case for
the Wegner type estimate (W). Note that, as $s\leq 1$,
$\varepsilon^sL(\log L)^\beta$ is small only when $\varepsilon L$ is
small. Finally, note that, as $\rho>1$, the factor
$(\varepsilon^sL(\log L)^\beta)^2$ is better i.e. smaller than
$(\varepsilon^sL^\rho)^2$, the square of the upper bound obtained by
the Wegner type estimate (W). This improvement is a consequence of
localization.\\
The estimate~\eqref{eq:8} is weaker than the Minami type estimate
found in~\cite{MR97d:82046,MR2360226,MR2290333,MR2505733} which gives
a bound on $\D\sum_{k\geq2}k\,
\P\left(\tr[\car_{[E-\varepsilon,E+\varepsilon]}
  (H_\omega(\Lambda_L))]\geq k\right)$. The estimate~\eqref{eq:8} is
nevertheless sufficient to repeat the analysis done
in~\cite{Ge-Kl:10,Kl:10a}. In particular, it is sufficient to obtain
the description of the eigenvalues of $H_\omega(\Lambda_L)$ in terms
of the ``approximated eigenvalues'' i.e. the eigenvalues of $H_\omega$
restricted to smaller cubes and to compute the law of those
approximated eigenvalues (see~\cite[Lemma 2.1, Theorem 1.15 and
1.16]{Ge-Kl:10},~\cite{Kl:10a}). \\
Let us now say a word how~\eqref{eq:8} can be used to apply the
analysis done in~\cite{Ge-Kl:10,Kl:10a} to the models $H^A_\omega$ and
$H^D_\omega$ studied in the introduction.\\
One checks that Theorem~\ref{thr:1} implies that, for any $s'\in(0,s)$
and $\eta>1$, there exists $L_{\eta,s'}>0$ s.t., for $E\in J$, $L\geq
L_{\eta,s'}$ and $\varepsilon\in\left(0,L^{-1/s'}\right)$, one has
\begin{equation}
  \label{eq:22}
  \sum_{k\geq2}\P\left(\tr[\car_{[E-\varepsilon,E+\varepsilon]}
    (H_\omega(\Lambda_L))]\geq k\right)\leq C\left(
    \varepsilon^{2s'}L^2+e^{-(\log L)^\eta/8}\right).
\end{equation}
The estimate~\eqref{eq:22} differs from the Minami estimate used
in~\cite{Ge-Kl:10,Kl:10a} in three ways. First,
in~\cite{Ge-Kl:10,Kl:10a}, it was assumed that $s=1$ and $\rho=1$ in
(W) and~\eqref{eq:22}. For $H^A_\omega$ and $H^D_\omega$,
in~\eqref{eq:22}, we have $\rho=1$ but only have $s'<1$ even though
arbitrary.  Second, we only have~\eqref{eq:22} under a smallness
condition on $\varepsilon$ (i.e. $\varepsilon\leq L^{-1/s'}$). Third,
in~\eqref{eq:22}, there is an additional error term $e^{-(\log
  L)^\eta/8}$. As already mentioned above, in the analysis performed
in~\cite{Ge-Kl:10,Kl:10a}, the Minami type estimate is used in two
ways: to control the occurrence of two eigenvalues in a small interval
for the operator restricted to a given box and to control the law of
approximated eigenvalues. For the first use, the crucial thing is that
if $L$ is the size of the box and $\varepsilon$ the size of the
interval, the bound in the Minami estimate should be very small
(see~\cite[Theorem 1.15 and 1.16]{Ge-Kl:10} and~\cite[section
3]{Kl:10a}). For the second use, the crucial thing is that the term
given by the Minami estimate should be smaller than than the main term
giving the law of these approximate eigenvalues which is of size
$\varepsilon\,L$ (see~\cite[Lemma 2.1]{Ge-Kl:10} and~\cite[Lemma
2.2]{Kl:10a}). In this application, the box size $L$ and $\varepsilon$
in~\eqref{eq:22} are related by a power law
i.e. $L=\varepsilon^{-\kappa}$ for some $\kappa<1$. So taking $s'$
sufficiently close to $1$ (which, for $H_\omega^A$ and $H_\omega^D$ is
possible as $s=1$ in (W)) guarantees that the condition
$\varepsilon\leq L^{-1/s'}$ is satisfied and, for $L$ large,
\begin{equation*}
  \varepsilon^{2s'}L^2+e^{-(\log L)^\eta/8}
  =o\left(\varepsilon L\right)
\end{equation*}
Before explaining the heuristics guiding the proofs of
Theorems~\ref{thr:1}, let us very briefly describe some consequences
for random operators. Essentially, all the conclusions described for
the models $H^A_\omega$ and $H^D_\omega$ in the introduction hold for
any general one dimensional random model satisfying the assumptions
(IAD), (W) and (Loc). As said in the introduction,
following~\cite{Ge-Kl:10,Kl:10a}, more results on the spectral
statistics can be obtained. As assumptions (IAD) and (Loc) have been
proved for many models e.g. the Poisson model
(see~\cite{MR2314108,MR2352267}), the Bernoulli Anderson model
(see~\cite{MR2180453}) or general Anderson models with non trivial
distributions (see~\cite{2011arxiv1105.0213G}), it remains to
understand Wegner type estimates (W) or replacements of such estimates
for those models.
\subsection{Inverse tunneling and the Minami type estimates}
\label{sec:inverse-tunneling-1}
To the best of our knowledge, up to the present work, the availability
of decorrelation estimates of the type~\eqref{eq:8} relied on the fact
that the single site potential was rank one
(\cite{MR97d:82046,MR2360226,MR2290333,MR2505733}) or had the
effective weight of a rank one potential as was shown
in~\cite{MR2663411} in the Lifshitz tails regime. In the present
paper, we exhibit a heuristic why such estimates should hold quite
generally and use it to develop a different approach. This approach
makes crucial use of localization to reduce the complexity of the
problem i.e. to study the random Hamiltonian restricted to some much
smaller cube. Such ideas were already used in~\cite{MR2775121} to
study spectral correlations at distinct energies.
We now turn to the heuristic we referred to earlier. The basic
mechanism at work in our heuristics is what we call ``inverse
tunneling''. Let us explain this and therefore, first recall
some facts on tunneling.\\
Fix $\ell\in\R$ and $q:\,[0,\ell]\to\R$ a real valued function bounded
by $Q>0$. On $[0,\ell]$, consider the Dirichlet eigenvalue problem
defined by the differential expression $-u''+qu$ i.e. the eigenvalue
problem
\begin{equation}
  \label{eq:24}
  -\frac{d^2}{dx^2}u(x)+q(x)u(x)=Eu(x),\quad u(0)=u(\ell)=0.
\end{equation}
Tunneling estimates can be described as follows. Assume that the
interval $[0,\ell]$ can be split into two intervals, say, $[0,\ell']$
and $[\ell',\ell]$, such that the Dirichlet eigenvalue problem for
each of those intervals share a common eigenvalue and such that the
associated eigenfunctions are ``very'' (typically exponentially) small
near $\ell'$ then the eigenvalue problem~\eqref{eq:24} has two
eigenvalues that are ``very'' close together. The closeness of the
eigenvalues and the smallness of the eigenfunctions are related; they
are in general measured in terms of some parameter e.g. a coupling
constant in front of the potential $q$, a semi-classical parameter in
front of the kinetic energy $-d^2/dx^2$ or the length of the interval
$\ell$ (see
e.g.~\cite{MR0347251,MR740094,Combes:1983su,MR707207}). The tunneling
effect is well illustrated by the double well problem (see
e.g.~\cite{MR581948}).\\
In the present paper, we discuss a converse to the above description
i.e. we assume that the eigenvalue problem~\eqref{eq:24} has two (or
more) close together eigenvalues, say, $0$ and $E$ small associated
respectively to $u$ and $v$. Let $\D r_u:=\sqrt{|u|^2+|u'|^2}$ and $\D
r_v:=\sqrt{|v|^2+|v'|^2}$ be the Pr{\"u}fer radii for $u$ and $v$ (see
e.g.~\cite{MR923320}). Then, either of two things happen:
\begin{enumerate}
\item no tunneling occurs i.e. $r_u\cdot r_v$ is small on the whole
  interval $[0,\ell]$. In this case, the eigenfunctions $u$ and $v$
  live in separate space regions and, thus, don't see each other.
\item tunneling occurs i.e. $r_u\cdot r_v$ becomes large in some region of
  space. In the connected components of such regions, $u$ and $v$ are
  roughly proportional. Thus, we show that it is possible to construct
  linear combinations of $u$ and $v$ that live in distinct space
  regions, that is, we undo the tunneling; these linear combinations
  are not true eigenfunctions anymore but they almost satisfy the
  eigenvalue equation as $E$ is close to $0$.
\end{enumerate}
In both cases, we construct quasi-modes that live in distinct space
regions (see section~\ref{sec:gener-splitt-result}). Thus, we
derive~\eqref{eq:8} using the Wegner type estimate (W) in each of
these regions (see section~\ref{sec:proof-theorem1}). This yields
Theorem~\ref{thr:1}.
\subsection{Universal estimates}
\label{sec:universal-estimates-1}
We now turn to deterministic estimates that are related to our
analysis of Minami estimates in one dimension. These estimates control
the minimal spacing between any two eigenvalues of a Schr{\"o}dinger
operator on $[0,\ell]$ (with Robin boundary conditions). By extension,
they also give an upper bound on the maximal number of eigenvalues a
Schr{\"o}dinger operator of $[0,\ell]$ can put inside an interval of size
$\varepsilon$. Though we do not know any reference for such estimates,
we are convinced that they are well known to the specialists.\\
For the sake of simplicity, assume $q:\,[0,\ell]\to\R$ is bounded. On
$[0,\ell]$, consider the operator $Hu=-u''+qu$ with self-adjoint Robin
boundary conditions at $0$ and $\ell$ (i.e. $u(0)\,\cos\alpha
+u'(0)\,\sin\alpha=0$). Then, one has
\begin{Th}
  \label{thr:3}
  Fix $J$ compact. There exists a constant $C>0$ (depending only on
  $\|q\|$ and $J$) such that, for $\ell\geq1$, if
  $\varepsilon\in(0,1)$ is such that $|\log\varepsilon|\geq C\ell$,
  then, for any $E\in J$, the interval $[E-\varepsilon,E+\varepsilon]$
  contains at most a single eigenvalue of $H$.
\end{Th}
\noindent One can generalize Theorem~\ref{thr:3} to
\begin{Th}
  \label{thr:6}
  Fix $\nu>2$ and $J$ compact. There exists $\ell_0>1$ and $C>0$
  (depending only on $\|q\|_\infty$ and $J$) such that, for
  $\ell\geq\ell_0$, if $\varepsilon\in(0,\ell^{-\nu})$ then, for $E\in
  J$, the number of eigenvalues of $H$ in the interval
  $[E-\varepsilon,E+\varepsilon]$ is bounded by
  $\max(1,C\ell/|\log\varepsilon|)$.
\end{Th}
\noindent These a-priori bounds prove that there is some repulsion
between the level for arbitrary Schr{\"o}dinger operators in dimension
one. For random systems in the localized phase, this repulsion takes
place at a length scale of size $e^{-C\ell}$; it is much smaller than
the typical level spacings that is of size $1/\ell$.\\
Similar results hold for discrete operators (see
e.g.~\cite{2011JPhA...44D5206R}).
\section{Inverse tunneling estimates}
\label{sec:inverse-tunneling}
\noindent Fix $\ell\in\R$ and $q:\,[0,\ell]\to\R$ a real valued
function bounded by $Q>0$. On $[0,\ell]$, consider the Sturm-Liouville
eigenvalue problem defined by
\begin{equation}
  \label{eq:25}
  (Hu)(x):=-\frac{d^2}{dx^2}u(x)+q(x)u(x)=Eu(x),\quad
  u(0)=0=u(\ell).
\end{equation}
\begin{Rem}
  \label{rem:5}
  Here, we use Dirichlet boundary conditions; the same analysis goes
  through with general Robin boundary conditions.
\end{Rem}
\noindent For $u$, a solution to~\eqref{eq:25}, define the Pr{\"u}fer
variables (see e.g.~\cite{MR923320}) by
\begin{equation*}
  r_u(x)\begin{pmatrix}\sin(\varphi_u(x))\\\cos(\varphi_u(x))
  \end{pmatrix}:=\begin{pmatrix} u(x)\\u'(x)\end{pmatrix},\quad
  r_u(x)>0,\quad \varphi_u(x)\in\R
\end{equation*}
Bu the Cauchy-Lipschitz Theorem, if $u$ does not vanish identically,
$r_u$ does not vanish. $\varphi_u$ is chosen so as to be
continuous. If $u$ is a solution to~\eqref{eq:25}, then we set
$\varphi_u(0)=0$ and $\varphi_u(\ell)=k\pi$ (for some
$k\in\N^*$). Rewritten in terms of the Pr{\"u}fer variables, the
eigenvalue equation in~\eqref{eq:25} becomes
\begin{gather}
  \label{eq:27}\varphi'_u(x)=1-(1+(q(x)-E))\sin^2(\varphi_u(x))\\
  \label{eq:28}\frac{r'_u(x)}{r_u(x)}=(1+(q(x)-E))\sin(\varphi_u(x))
  \cos(\varphi_u(x)).
\end{gather}
Let us now compare eigenfunctions associated to close by eigenvalues.
\subsection{General estimates for eigenfunctions associated to close
  by eigenvalues }
\label{sec:gener-estim-eigenf}
Consider now $u$ and $v$ two normalized eigenfunctions of the
Sturm-Liouville problem~\eqref{eq:25} associated to two consecutive
eigenvalues, say, $0$ and $E$. We assume $0<E\ll 1$. Sturm's
oscillation theorem then guarantees that $\varphi_u(x)<\varphi_v(x)$
for $x\in(0,\ell)$ and $\varphi_v(\ell)=\varphi_u(\ell)+\pi$ (see
e.g.~\cite{MR923320}). Define
\begin{equation}
  \label{eq:29}
  \delta\varphi(x)=\varphi_v(x)-\varphi_u(x).
\end{equation}
Thus, $\delta\varphi(0)=0$, $\delta\varphi(\ell)=\pi$ and
$\delta\varphi(x)\in(0,\pi)$ for $x\in(0,\ell)$.\\
The function $\delta\varphi$ satisfies the following differential
equation
\begin{equation}
  \label{eq:31}
  \begin{split}
    (\delta\varphi)'(x)&=(1+q(x))[\sin^2(\varphi_v(x))-\sin^2(\varphi_u(x))]
    -E\sin^2(\varphi_v(x))
    \\&=(1+q(x))\sin(\delta\varphi(x))\sin(2\varphi_v(x)-\delta\varphi(x))
    -E\sin^2(\varphi_v(x)).
  \end{split}
\end{equation}
The first property we prove is that, on intervals where
$\sin(\delta\varphi(x))$ is small, $r_u$ and $r_v$ are essentially
proportional to each other, that is,
\begin{Le}
  \label{le:9}
  Fix $\varepsilon>0$. Assume that, for $x\in[x_-,x_+]$, one has
  $\sin(\delta\varphi(x))\leq\varepsilon$. Then, there exists
  $\lambda>0$ such that, for $x\in[x_-,x_+]$, one has
  \begin{equation}
    \label{eq:41}
    e^{-[(Q+1)\varepsilon+E]\ell}\leq
    \frac{r_v(x)}{r_u(x)}\frac1\lambda\leq
    e^{[(Q+1)\varepsilon+E]\ell}.
  \end{equation}
\end{Le}
\begin{proof}
  Comparing~\eqref{eq:28} for $u$ and $v$ yields
  \begin{equation}
    \label{eq:37}
    \left[\log\left(\frac{r_v(x)}{r_u(x)}\right)\right]'=
    (1+q(x))\sin(\delta\varphi(x))\cos(2\varphi_v(x)-\delta\varphi(x))
    -E\sin(2\varphi_v(x))
  \end{equation}
  As, for $x\in[x_-,x_+]$ one has $0\leq
  \sin(\delta\varphi(x))\leq\varepsilon$,~\eqref{eq:37} yields, for
  $x\in[x_-,x_+]$,
  \begin{equation*}
    \left|\left[\log\left(\frac{r_v(x)}{r_u(x)}\right)\right]'\right|\leq
    (1+Q)\varepsilon+E.
  \end{equation*}
  Integrating this equation, for $(x,y)\in[x_-,x_+]^2$, one obtains
  \begin{equation}
    \label{eq:38}
    e^{-[(Q+1)\varepsilon+E](y-x)}\frac{r_v(y)}{r_u(y)}\leq
    \frac{r_v(x)}{r_u(x)}\leq
    e^{[(Q+1)\varepsilon+E](y-x)}\frac{r_v(y)}{r_u(y)}.
  \end{equation}
  This in particular immediately yields~\eqref{eq:41} and completes
  the proof of Lemma~\ref{le:9}.
\end{proof}
Next we prove that the Wronskian of $u$ and $v$ does not vary much on
intervals over which $\sin(\delta\varphi(x))$ is ``large'', that is,
\begin{Le}
  \label{le:11}
  Fix $\varepsilon>0$ such that $E<\varepsilon<1$.  Assume that
  \begin{itemize}
  \item for $x\in[x_-,x_+]$, one has
    $\sin(\delta\varphi(x))\geq\varepsilon$;
  \item $\sin(\delta\varphi(x_\pm))=\varepsilon$.
  \end{itemize}
  Then, for $x\in[x_-,x_+]$, one has $w(v,u)(x)>0$ and
  \begin{equation*}
    \max_{x\in[x_-,x_+]}\left[1-\frac{w(v,u)(x)}
      {\D\max_{x\in[x_-,x_+]}w(v,u)(x)}\right]\leq
    \frac{E}{\varepsilon}(x_+-x_-)\leq
    \frac{E\ell}{\varepsilon}.
  \end{equation*}
\end{Le}
\begin{proof}
  Consider the Wronskian of $v$ and $u$, that is,
  $w(v,u)(x)=u'(x)v(x)-v'(x)u(x)$. As $u$ and $v$ are eigenfunctions
  for the same Sturm-Liouville problem for the eigenvalues $0$ and
  $E$, $w(v,u)$ satisfies the equation $[w(v,u)]'=Euv$. Thus, for
  $(x,y)\in[0,\ell]^2$, one has
  \begin{gather}
    \label{eq:32}
    w(v,u)(x)=r_u(x)r_v(x)\sin(\delta\varphi(x)),\\
    \intertext{and} \label{eq:33}
    \begin{split}
      r_u(x)r_v(x)\sin(\delta\varphi(x))&-r_u(y)r_v(y)\sin(\delta\varphi(y))
      \\&=E\int_y^xr_u(t)r_v(t)\sin(\varphi_u(t))\sin(\varphi_v(t))dt.
    \end{split}
  \end{gather}
  The positivity of $w(v,u)$ is a direct consequence of~\eqref{eq:32}
  and the assumption on $[x_-,x_+]$. \\
  As for $x\in[x_-,x_+]$, one has
  $\sin(\delta\varphi(x))\geq\varepsilon$, using~\eqref{eq:32}, for
  $(x,y)\in[x_-,x_+]^2$, one can rewrite~\eqref{eq:33} as
  \begin{equation*}
    w(v,u)(y)-w(v,u)(x)=\frac{E}\varepsilon
    \int_x^yw(v,u)(t)g(t)dt\quad\text{ where }\quad\sup_{t\in[x,y]}|g(t)|\leq 1.
  \end{equation*}
  Thus, for $y$ such that $\D
  w(v,u)(y)=\max_{x\in[x_-,x_+]}w(v,u)(x)$, we obtain
  \begin{equation*}
    0\leq 1-\frac{w(v,u)(x)}{\D\max_{x\in[x_-,x_+]}w(v,u)(x)}=\frac{E}\varepsilon
    \int_x^y\frac{w(v,u)(t)}{\D\max_{x\in[x_-,x_+]}w(v,u)(x)}g(t)dt
    \leq\frac{E}{\varepsilon}|y-x|
  \end{equation*}
  This completes the proof of Lemma~\ref{le:11}.
\end{proof}
\noindent Next, we give another result showing that the Wronskian of
$u$ and $v$ does not vary much on intervals over which
$\sin(\delta\varphi(x))$ is ``large'', namely,
\begin{Le}
  \label{le:10}
  Fix $\varepsilon>0$ such that $E<\varepsilon<1$. Assume that
  \begin{itemize}
  \item for $x\in[x_-,x_+]$, one has
    $\sin(\delta\varphi(x))\geq\varepsilon$;
  \item $\sin(\delta\varphi(x_\pm))=\varepsilon$.
  \end{itemize}
  Then, for any $a>1$,
  \begin{itemize}
  \item either $r_u(x_+)r_v(x_+)+r_u(x_-)r_v(x_-)\leq
    2a\,\ell(E/\varepsilon)^2$,
  \item or
    \begin{equation*}
      1-\frac{1}{1+a}
      \leq\frac{r_u(x_-)r_v(x_-)}{r_u(x_+)r_v(x_+)}\leq
      1+\frac{1}{a}.
    \end{equation*}
  \end{itemize}
\end{Le}
\begin{proof}
  Recall that the system $(u,v)$ is orthonormal in $L^2([0,\ell])$;
  thus, one has
  \begin{gather}
    \label{eq:35}
    \int_0^lr^2_u(x)\sin^2(\varphi_u(x))dx=1=
    \int_0^lr^2_v(x)\sin^2(\varphi_v(x))dx,\\
    \label{eq:36}
    \int_0^lr_u(x)r_v(x)\sin(\varphi_u(x))\sin(\varphi_v(x))dx=0.
  \end{gather}
  As $w(v,u)(0)=w(v,u)(\ell)=0$ $\D$, this and~\eqref{eq:33} implies
  that
  \begin{equation}
    \label{eq:34}
    0<\max_{x\in[0,\ell]}\,r_u(x)r_v(x)\sin(\delta\varphi(x))\leq E.
  \end{equation}
  As for $x\in[x_-,x_+]$, one has
  $\sin(\delta\varphi(x))\geq\varepsilon/2$, one obtains
  \begin{equation}
    \label{eq:51}
    0<\max_{x\in[x_-,x_+]}\,r_u(x)r_v(x)\leq 2E/\varepsilon.
  \end{equation}
  Inserting this estimate into~\eqref{eq:33} for $(x,y)=(x_-,x_+)$ and
  using the fact that $\sin(\delta\varphi(x_-))=\varepsilon
  =\sin(\delta\varphi(x_+))$, one obtains
  \begin{equation*}
    |r_u(x_-)r_v(x_-)-r_u(x_+)r_v(x_+)|\leq
    2\ell(E/\varepsilon)^2.
  \end{equation*}
  This implies the alternative asserted by Lemma~\ref{le:10}.
\end{proof}
\subsection{An inverse "splitting" result}
\label{sec:gener-splitt-result}
We prove
\begin{Th}
  \label{thr:7}
  Fix $S>0$ arbitrary and $J\subset\R$ a compact interval. There
  exists $\varepsilon_0>0$ and $\ell_0>0$ (depending only on
  $\|q\|_\infty$, $J$ and $S$) such that, for $\ell\geq\ell_0$ and
  $0<\varepsilon\ell^4\leq\varepsilon_0$, for $E\in J$, if the
  operator $H$ defined in~\eqref{eq:25} has two eigenvalues in
  $[E-\varepsilon,E+\varepsilon]$, then there exists two points $x_+$
  and $x_-$ in the lattice segment $\varepsilon_0\Z\cap[0,\ell]$
  satisfying $S< x_+-x_-<2S$ such that, if $H_-$, resp. $H_+$, denotes
  the second order differential operator $H$ defined by~\eqref{eq:24}
  and Dirichlet boundary conditions on $[0,x_-]$, resp. on
  $[x_+,\ell]$, then $H_-$ and $H_+$ each have an eigenvalue in the
  interval $[E-\varepsilon\ell^4/\varepsilon_0,
  E+\varepsilon\ell^4/\varepsilon_0]$.
\end{Th}
\noindent Theorem~\ref{thr:7} is a consequence of
Propositions~\ref{pro:3} and~\ref{pro:1} that are respectively proved
in sections~\ref{sec:when-there-is} and~\ref{sec:when-there-no}. Let
us now explain the ideas guiding the proof of Theorem~\ref{thr:7}.\\
Up to a shift in energy and potential $q$, we may assume that $E=0$
and that, changing the notations, the eigenvalues considered in
Theorem~\ref{thr:7} are $0$ and $E>0$. All the estimates we will prove
only depend on $\|q\|_\infty$ in this new setting, thus, only depend
on $\|q\|_\infty$ and $J$ in the old setting. Note that, in the new
notations we have $E\leq \varepsilon$.\\
Let $u$ and $v$ be the eigenfunctions associated respectively to $0$
and $E$. The goal is then to prove that we can find two independent
linear combinations of $u$ and $v$ such that
\begin{itemize}
\item they vanish at two points, say, $x_-$ and $x_+$ satisfying the
  statement of Theorem~\ref{thr:7},
\item in each of these intervals $[0,x_-]$ and $[x_+,\ell]$, the
  masses of the combinations are of size of order $\ell^{-\alpha}$
  (for some $\alpha>0$).
\end{itemize}
Therefore, we consider two cases:
\begin{enumerate}
\item if $r_u\cdot r_v$ becomes ``large'' over $[0,\ell]$ which we dub the
  ``tunneling case''.
\item if $r_u\cdot r_v$ stays ``small'' over $[0,\ell]$ which we dub the
  ``non tunneling case''.
\end{enumerate}
In the first case, $u$ and $v$ put mass at the same locations in
$[0,\ell]$. This is typically what happens in a tunneling situation
(see e.g.~\cite{MR0347251,MR740094,Combes:1983su,MR707207}). In this
case, there is a strong ``interaction'' between $u$ and $v$ and the
estimates obtained in section~\ref{sec:inverse-tunneling} enable us to
show that $u$ and $v$ are quite similar up to a phase change. Although
they are linearly independent (as they are eigenfunctions associated
to distinct eigenvalues of the same self-adjoint operator), they are
similar in the sense that $r_u$ and $r_v$ are similar (see
Lemma~\ref{le:9}). Their orthogonality comes mainly from the phase
difference. We analyze this phase difference to prove that the claims
of Theorem~\ref{thr:7} hold in this case.\\
In the second case, $u$ and $v$ live ``independent lives''; they are
of course orthogonal but $|u|$ and $|v|$ (actually, $r_u$ and $r_v$
too) are also almost orthogonal. So, $u$ and $v$ roughly live on
disjoint sets; this makes it quite simple to construct the functions
whose existence is claimed in Theorem~\ref{thr:7}: one just needs to
restrict $u$ and $v$ to their ``essential supports''.
\subsubsection{When there is tunneling}
\label{sec:when-there-is}
The case when there is tunneling can be described by the fact that the
function $u$ and $v$ are ``large'' at the same location or
equivalently by the fact that $r_u\cdot r_v$ becomes ``large'' at some
point of the interval $[0,\ell]$. Clearly, as $u$ and $v$ are
normalized, $r_u$ and $r_v$ need each to be at most only of size
$1/\sqrt{\ell}$. So one can say that $r_u\cdot r_v$ becomes ``large'' if
and only if $r_u\cdot r_v\gtrsim\ell^{-1}$ somewhere in $[0,\ell]$.\\
We prove
\begin{Pro}
  \label{pro:3}
  Fix $S>0$ arbitrary. There exists $\eta_0>0$ (depending only on $S$
  and $\|q\|_\infty$) such that, for $\eta\in(0,\eta_0)$ and $\ell$
  sufficiently large (depending only on $\eta$, $S$ and
  $\|q\|_\infty$), if $u$ and $v$ are as in
  section~\ref{sec:inverse-tunneling}, that is, eigenfunctions of $H$
  associated respectively to the eigenvalues $0$ and $E$, and, assume
  that $E\ell^4\leq\eta^4$ and that one has
  \begin{equation}
    \label{eq:45}
    \exists\,x_0\in[0,\ell],\quad r_u(x_0)\cdot r_v(x_0)\geq\frac{\eta}{\ell},
  \end{equation}
  then, there exists two points $x_+$ and $x_-$ in the lattice segment
  $\eta\Z\cap[0,\ell]$ satisfying
  \begin{equation}
    \label{eq:46}
    |\log(E\ell^2)|/C<x_-<x_+<\ell-|\log(E\ell^2)|/C\quad\text{and}\quad
    S< x_+-x_-<2S
  \end{equation}
  such that, if $H_-$ (resp. $H_+$) denotes the second order
  differential operator $H$ defined by~\eqref{eq:25} and Dirichlet
  boundary conditions on $[0,x_-]$ (resp. on $[x_+,\ell]$), then $H_-$
  and $H_+$ have an eigenvalue in the interval
  $[-E\ell^4\eta^{-4},E\ell^4\eta^{-4}]$.
\end{Pro}
\begin{proof} We keep the notations of
  section~\ref{sec:gener-estim-eigenf}. By~\eqref{eq:30} for $u$ and
  $v$,~\eqref{eq:45} implies that there exists $C>0$ (depending only
  on $\|q\|_\infty$) such that
  \begin{equation}
    \label{eq:48}
    \forall x\in[x_0-1,x_0+1]\cap[0,\ell]\quad
    r_u(x)\cdot r_v(x)\geq\frac{\eta}{C\ell},
  \end{equation}
  Note that, by~\eqref{eq:45} and~\eqref{eq:34}, one has
  \begin{equation}
    \label{eq:14}
    \forall x\in[x_0-1,x_0+1]\cap[0,\ell],\quad  \sin(\delta\varphi(x))\lesssim
    E\ell/\eta.
  \end{equation}
  For the sake of definiteness, we assume that
  \begin{equation}
    \label{eq:47}
    \forall x\in[x_0-1,x_0+1]\cap[0,\ell],\quad 0\leq
    \delta\varphi(x)\lesssim E\ell/\eta,
  \end{equation}
  the case $0\leq\pi-\delta\varphi(x)\lesssim E\ell/\eta$ being dealt
  with in the same way. \\
  As $u$ and $v$ are normalized and orthogonal to each other, one
  proves
  \begin{Le}
    \label{le:14}
    There exists $C>0$ (depending only on $\|q\|_\infty$) and
    $x_2\in[0,\ell]$ such that, for $x\in[x_2-1,x_2+1]\cap[0,\ell]$,
    one has
    \begin{equation}
      \label{eq:49}
      r_u(x)\cdot r_v(x)\geq\frac{\eta}{C\ell^2}\quad\text{and}\quad
      0\leq\pi-\delta\varphi(x)\lesssim E\ell^2/\eta.
    \end{equation}
  \end{Le}
  \begin{Rem}
    \label{rem:7} When $0\leq\pi-\delta\varphi(x)\lesssim E\ell/\eta$
    on $[x_0-1,x_0+1]\cap[0,\ell]$, in~\eqref{eq:49}, the statement
    $0\leq\pi-\delta\varphi(x)\lesssim E\ell^2/\eta$ is replaced with
    $0\leq \delta\varphi(x)\lesssim E\ell^2/\eta$.
  \end{Rem}
  \begin{proof}
    Indeed, by~\eqref{eq:48} and~\eqref{eq:47}, one has
    \begin{multline}
      \label{eq:57}
      \left|\int_{[x_0-1,x_0+1]\cap[0,\ell]}r_u(x)r_v(x)
        \sin^2(\varphi_u(x))dx\right.
      \\\left.+\int_{[0,\ell]\setminus[x_0-1,x_0+1]}
        r_u(x)r_v(x)\sin(\varphi_u(x))\sin(\varphi_v(x))dx\right|\lesssim
      \frac{E\ell}{\eta}.
    \end{multline}
    Hence, by~\eqref{eq:26} in Lemma~\ref{le:13} and~\eqref{eq:48}, as
    $E\ell^4\leq\eta_0$, we get that, for some $C>0$ (depending only
    on $\|q\|_\infty$), one has
    \begin{equation}
      \label{eq:50}
      \begin{split}
        \int_{[0,\ell]\setminus[x_0-1,x_0+1]}
        r_u(x)r_v(x)\sin(\varphi_u(x))\sin(\varphi_v(x))dx&\leq
        -\frac{\eta}{C\ell}\left(1-\frac{C\eta^2}{\ell^2}\right)
        \\&\lesssim-\frac{\eta}{\ell}
      \end{split}
    \end{equation}  
    for $\ell$ sufficiently large.\\
    Write
    \begin{equation}
      \label{eq:40}
      \begin{split}
        &\int_{[0,\ell]\setminus[x_0-1,x_0+1]}
        r_u(x)r_v(x)\sin(\varphi_u(x))\sin(\varphi_v(x))dx\\&\hskip1cm=
        \int_{\substack{x\in[0,\ell]\setminus[x_0-1,x_0+1]\\r_u(x)r_v(x)\leq
            \eta/\ell^2}}
        r_u(x)r_v(x)\sin(\varphi_u(x))\sin(\varphi_v(x))dx\\&\hskip2cm+
        \int_{\substack{x\in[0,\ell]\setminus[x_0-1,x_0+1]\\r_u(x)r_v(x)>
            \eta/\ell^2}}
        r_u(x)r_v(x)\sin(\varphi_u(x))\sin(\varphi_v(x))dx
      \end{split}
    \end{equation}  
    By~\eqref{eq:34}, on the set $\D\{x\in[0,\ell];\ r_u(x)r_v(x)>
    \eta/\ell^2\}$, one has $\sin(\delta\varphi(x))\leq
    E\ell^2/\eta$. Thus, as $E\ell^4\leq\eta$,~\eqref{eq:40} yields
    \begin{equation}
      \label{eq:39}
      \begin{split}
        \int_{\substack{x\in[0,\ell]\setminus[x_0-1,x_0+1]\\
            r_u(x)r_v(x)> \eta/\ell^2\\\sin(\delta\varphi(x))\leq
            E\ell^2/\eta}}
        r_u(x)r_v(x)\sin^2(\varphi_u(x))\cos(\delta\varphi(x))dx&\leq
        -\frac{\eta}{2C\ell}\left(1-\frac{C\eta^2}{\ell^2}\right)
        \\&\lesssim-\frac{\eta}{\ell}.
      \end{split}
    \end{equation}  
    This and~\eqref{eq:30} then proves Lemma~\ref{le:14}.
  \end{proof}
  \noindent Clearly, by~\eqref{eq:47} and~\eqref{eq:49}, one has
  $[x_0-1,x_0+1]\cap[x_2-1,x_2+1]=\emptyset$. For the sake of
  definiteness, assume that $x_0+1<x_2-1$. By~\eqref{eq:47}
  and~\eqref{eq:50}, as $x\mapsto\delta\varphi(x)$ is continuous,
  there exists $x_0+1<x_1<x_2-1$ such that
  $\sin(\delta\varphi(x_1))=1$, that is,
  $\delta\varphi(x_1)=\pi/2$.\\
  Fix now $\varepsilon=\eta/(C\ell^2)$. By~\eqref{eq:14}
  and~\eqref{eq:49}, there exists two intervals $[x_0^-,x_0^+]$ and
  $[x_2^-,x_2^+]$ such that,
  \begin{itemize}
  \item
    $[x_0-1,x_0+1]\cap[0,\ell]\subset[x_0^-,x_0^+]\subset[0,\ell]$;
  \item
    $[x_2-1,x_2+1]\cap[0,\ell]\subset[x_2^-,x_2^+]\subset[0,\ell]$;
  \item for $x\in[x_0^-,x_0^+]\cup[x_2^-,x_2^+]$, one has
    $\sin(\delta\varphi(x))\leq\varepsilon$;
  \item
    $\sin(\delta\varphi(x_0^\pm))=\sin(\delta\varphi(x_2^\pm))=\varepsilon$.
  \end{itemize}
  As $x_0+1<x_2-1$, one has $0<x_0^-<x_0^+<x_2^-<x_2^+<\ell$. This
  also implies that $[x_0,x_0+1]\subset[0,\ell]$ and
  $[x_2-1,x_2]\subset[0,\ell]$. Moreover, there exists a segment
  $[x_1^-,x_1^+]$ such that
  \begin{itemize}
  \item $x_1\in[x_1^-,x_1^+]\subset[x_0^+,x_2^-]\subset[0,\ell]$ ,
  \item for $x\in[x_1^-,x_1^+]$, one has
    $\sin(\delta\varphi(x))\geq\varepsilon$ and
    $\sin(\delta\varphi(x_1^\pm))=\varepsilon$.
  \end{itemize}
  As $\sin(\delta\varphi(x_1))=1$, by Lemma~\ref{le:15}, for some
  $C>0$ (depending only on $S$ and $\|q\|_\infty$), one has
  \begin{equation}
    \label{eq:15}
    \min_{x\in[x_1-2S,x_1+2S]}\sin(\delta\varphi(x))\geq \frac1C.
  \end{equation}
  Thus, for $\ell$ sufficiently large, as $\varepsilon<1/C$, one has
  $[x_1-2S,x_1+2S]\in[x_1^-,x_1^+]$.\\
  By Lemma~\ref{le:15}, we know that
  \begin{equation}
    \label{eq:52}
    C^{-1}|\log(E\ell^2)|\leq x_0^+-x_0^-\quad \text{and}\quad
    C^{-1}|\log(E\ell^3)|\leq x_2^+-x_2^-
  \end{equation}
  We apply Lemma~\ref{le:9} to $[x_0^-,x_0^+]$ and
  $[x_2^-,x_2^+]$. Hence, for $\ell$ sufficiently large,~\eqref{eq:41}
  implies that there exists $\lambda_0>0$ and $\lambda_2>0$ such that,
  for $i\in\{0,2\}$, one has
  \begin{equation}
    \label{eq:56}
    \frac{\lambda_i}{1+C\eta_0/\ell} \leq\min_{x\in[x_i^-,x_i^+]}
    \left(\frac{r_u(x)}{r_v(x)}\right)\leq\max_{x\in[x_i^-,x_i^+]}
    \left(\frac{r_u(x)}{r_v(x)}\right)\leq
    \left[1+C\eta_0/\ell\right]\lambda_i.
  \end{equation}
  Moreover, as $r_u$ and $r_v$ are bounded by a constant depending
  only on $\|q\|_\infty$, by~\eqref{eq:52}, \eqref{eq:48}
  and~\eqref{eq:49}, one has $\D \frac\eta{C\ell^2}\leq
  \lambda_0,\lambda_2\leq\frac{C\ell^2}{\eta}$.\\
  By Lemma~\ref{le:11}, on $[x_1^-,x_1^+]$, one has
  \begin{equation}
    \label{eq:54}
    w(u,v)(x)=M\left(1+O\left(
        \frac{E\ell}{\varepsilon}\right)\right)
    \quad\text{where}\quad
    M:=\max_{x\in[x_1^-,x_1^+]}w(u,v)(x).
  \end{equation}
  We prove
  \begin{Le}
    \label{le:16}
    There exists $\eta_0>0$ (depending only on $\|q\|_\infty$) such
    that, for $\eta\in(0,\eta_0)$, there exists $(k_-,k_+)\in\N^2$
    such that
    \begin{enumerate}
    \item $\frac{2S}3<x_1-k_-\eta<\frac{3S}4$ and
      $\frac{2S}3<k_+\eta-x_1<\frac{3S}4$;
    \item there exists $\lambda_\pm\in\R$ s.t. for
      $\bullet\in\{+,-\}$, one has
      \begin{itemize}
      \item either $u(k_\bullet\eta)=\lambda_\bullet v(k_\bullet\eta)$
        and
        \begin{itemize}
        \item $|\lambda_--\lambda_0|\geq \eta_0\eta$,
        \item $|\lambda_++\lambda_2|\geq \eta_0\eta$.
        \end{itemize}
      \item or $v(k_\bullet\eta)=\lambda_\bullet u(k_\bullet\eta)$ and
        \begin{itemize}
        \item $|\lambda_--1/\lambda_0|\geq \eta_0\eta$,
        \item $|\lambda_++1/\lambda_2|\geq \eta_0\eta$.
        \end{itemize}
      \end{itemize}
    \end{enumerate}
  \end{Le}
  \begin{proof}
    The proofs of the existence of $k_-$ and $k_+$ being the same up
    to obvious modifications, we only give the details for $k_-$.\\
    By Lemma~\ref{le:17}, there exists $\eta_0>0$ (depending only on
    $\|q\|_\infty$) such that, for $\eta\in(0,\eta_0)$,
    $|\sin(\varphi_u(x))|$ and $|\sin(\varphi_v(x))|$ can stay smaller
    than $\eta$ only on intervals of length less than
    $\eta/\eta_0$. Thus, there exits $\eta_0>0$ such that, for
    $\eta\in(0,\eta_0)$, one can find an integer $k$ such that
    \begin{equation}
      \label{eq:55}
      \frac{2S}3<x_1-k\eta<\frac{3S}4\quad\text{and}\quad 
      \begin{aligned}
        |\sin(\varphi_u(x))|\geq\eta\\
        |\sin(\varphi_v(x))|\geq\eta
      \end{aligned}
      \quad\text{ for }x\in[(k-1)\eta,(k+1)\eta].
    \end{equation}
    This, in particular, implies that $u(x)\not=0\not=v(x)$ for
    $x\in[(k-1)\eta,(k+1)\eta]$. Note also that, by~\eqref{eq:52}
    and~\eqref{eq:55}, for $\ell$ large, one has
    $k\eta\in[x_1-2S,x_1+2S]\subset[x_1^-,x_1^+]$. \\
    To fix ideas, assume moreover that $r_u(k\eta)\geq r_v(k\eta)$;
    the reverse case is treated similarly interchanging $u$ and $v$,
    and, $\lambda_{0}$ and $1/\lambda_{0}$. This in particular implies
    that, for some constant $C>0$ (depending only on $\|q\|_\infty$,
    see equation~\eqref{eq:37}), one has
    \begin{equation}
      \label{eq:73}
      r_u(x)\geq r_v(x)e^{-C\eta}\quad\text{for}\quad
      x\in[(k-1)\eta,(k+1)\eta].
    \end{equation}
    Assume that the first point of (2) in Lemma~\ref{le:16} does not
    hold i.e. assume now that
    \begin{equation}
      \label{eq:53}
      \exists\,\lambda\in\left[\lambda_{0}-\eta_0\eta,
        \lambda_{0}+\eta_0\eta\right]
      \quad\text{such that}\quad u(k\eta)=\lambda v(k\eta).
    \end{equation}
    As $v^2\cdot(u/v)'=w(u,v)$, we compute
    \begin{equation*}
      \begin{split}
        \frac{u(k\eta+\eta)}{v(k\eta+\eta)}&=
        \frac{u(k\eta)}{v(k\eta)}+\eta\int_0^1\frac{w(u,v)(k\eta+\eta
          t)}{v^2(k\eta+\eta t)}dt\\&=
        \lambda+\eta\int_0^1\frac{w(u,v)(k\eta+\eta t)}{v^2(k\eta+\eta
          t)}dt.
      \end{split}
    \end{equation*}
    Using successively
    \begin{itemize}
    \item the uniform estimate on the growth rate of $r_v$ given by
      equation~\eqref{eq:28},
    \item the estimate~\eqref{eq:54} on the Wronskian $w(u,v)$,
    \item the assumption $r_u(k\eta)\leq r_v(k\eta)$,
    \item the bound~\eqref{eq:15},
    \item and, presumably, a reduction of the value $\eta_0$,
    \end{itemize}
    we compute
    \begin{equation*}
      \begin{split}
        \int_0^1\frac{w(u,v)(k\eta+\eta t)}{v^2(k\eta+\eta
          t)}dt&\geq\frac1{Cr_v^2(k\eta)}\int_0^1w(u,v)(k\eta+\eta
        t)dt\geq\frac{M}{C r^2_v(k\eta)}\\&\geq\frac{M}{C
          r_v(k\eta)r_u(k\eta)}\geq\frac1C \frac{M}{
          w(u,v)(k\eta)}\geq\frac1C\geq2\eta_0.
      \end{split}
    \end{equation*}
    Thus, one has
    \begin{equation*}
      u(k\eta+\eta)=(\lambda+\delta\lambda)v(k\eta+\eta)\quad\text{with}
      \quad \lambda+\delta\lambda-\lambda_{0}\geq
      \delta\lambda-|\lambda-\lambda_{0}|\geq\eta\eta_0
    \end{equation*}
    and we set $k_-=k+1$.\\
    If~\eqref{eq:53} does not hold, it suffices to set $k_-=k$.\\
    This completes the proof of Lemma~\ref{le:16}.
  \end{proof}
  \noindent To complete the proof of Proposition~\ref{pro:1}, we check
  the assertion about $H_-$; the one about $H_+$ is checked likewise
  except for the fact that $\ell$ has to be replaced by $\ell^2$,
  compare~\eqref{eq:49} in Lemma~\ref{le:14} with~\eqref{eq:48}.\\
  The proof of Proposition~\ref{pro:1} now depends on which of the
  alternatives of Lemma~\ref{le:16} is realized. First, assume that,
  in Lemma~\ref{le:16}, it is the function $u-\lambda_- v$ that
  vanishes at $x_-=k_-\eta$. So, the function $u-\lambda_- v$
  satisfies Dirichlet boundary conditions on the interval $[0,x_-]$.
  One computes
  \begin{equation*}
    \|(H_--E)(u-\lambda_-v)\|_{L^2([0,x_-])}=E\|u\|_{L^2([0,x_-])}\leq E.
  \end{equation*}
  Moreover, by the defining property of $[x_0^-,x_0^+]$ and
  Lemma~\ref{le:9}, as $|\lambda_--\lambda_0|\geq\eta_0\eta$,
  using~\eqref{eq:56}, for $x\in[x_0^-,x_0^+]$, one has
  \begin{equation*}
    \begin{split}
      u(x)-\lambda_-
      v(x)&=r_u(x)\sin(\varphi_u(x))-\lambda_-r_v(x)\sin(\varphi_v(x))
      \\&=[r_u(x)-\lambda_-r_v(x)]\sin(\varphi_v(x))+O(E|u'(x)|)
      \\&=[\lambda_0-\lambda_-]v(x)
      +O(E|u'(x)|)+O(E^2|u(x)|)+O(\eta_0/\ell|v(x)|)
    \end{split}
  \end{equation*}
  Possibly reducing $\eta_0$, one then computes
  \begin{equation*}
    \|u-\lambda_-v\|_{L^2([0,x_-])}\geq \eta_0(\eta-1/\ell)
    \|v\|_{L^2([x_0-1,x_0+1])}- CE\ell\geq \eta_0\eta^3\ell^{-2}-
    CE\ell.
  \end{equation*}
  Thus, we know that $H_-$ has an eigenvalue at distance at most
  $E\ell^2/(2\eta_0\eta^3)$ from $E$ if
  $\eta_0\eta^3\ell^{-2}\gtrsim E>0$.\\
  When, in Lemma~\ref{le:16}, it is the function $v-\lambda_- u$ that
  vanishes at $x_-=k_-\eta$, one computes
  $\|H_-(v-\lambda_-u)\|_{L^2([0,x_-])}$ and the remaining part of the
  proof is unchanged.\\
  This completes the proof of Proposition~\ref{pro:3}.
\end{proof}
\noindent When we use Theorem~\ref{thr:7} to derive
Theorem~\ref{thr:1}, it will be of importance to have two points $x_-$
and $x_+$ that are well separated from each other. But, minor changes
in the proof of Proposition~\ref{pro:3} also yield the following
result
\begin{Pro}
  \label{pro:4}
  Fix $S>0$ arbitrary. There exists $\eta_0>0$ such that, for
  $\eta\in(0,\eta_0)$ and $\ell$ sufficiently large (depending only on
  $\eta$, $S$ and $\|q\|_\infty$), if $u$ and $v$ are as
  section~\ref{sec:inverse-tunneling} and such that~\eqref{eq:45} is
  satisfied, then, there exists a points $\overline{x}$ in the lattice
  $\eta\Z$ satisfying
  \begin{equation*}
    |\log\eta|/C<\overline{x}<\ell-|\log\eta|/C
  \end{equation*}
  such that, if $H_-$ (resp. $H_+$) denotes the second order
  differential operator $H$ defined by~\eqref{eq:25} and Dirichlet
  boundary conditions on $[0,\overline{x}]$ (resp. on
  $[\overline{x},\ell]$), then $H_-$ and $H_+$ have an eigenvalue in
  the interval $[-E\ell^4\eta^{-4},E\ell^4\eta^{-4}]$.
\end{Pro}
\subsubsection{When there is no tunneling}
\label{sec:when-there-no}
The case when there is no tunneling can be described by the fact that
both function $u$ and $v$ are ``large'' only at distinct location or
equivalently by the fact that $r_u\cdot r_v$ stays small all over the
interval $[0,\ell]$. Clearly, as $u$ and $v$ are normalized, $r_u$ and
$r_v$ need each to be at most only of size $1/\sqrt{\ell}$. So one can
say that $r_u\cdot r_v$ stays small if and only if $r_u\cdot r_v\ll \ell^{-1}$
all over $[0,\ell]$. We prove
\begin{Pro}
  \label{pro:1}
  Fix $S>0$ arbitrary. There exists $\eta_0>0$ (depending only on
  $\|q\|_\infty$) such that, for $\eta\in(0,\eta_0)$ and $\ell$
  sufficiently large (depending only on $\eta$, $S$ and
  $\|q\|_\infty$), if $u$ and $v$ are as
  section~\ref{sec:inverse-tunneling} that is, eigenfunctions of $H$
  associated respectively to the eigenvalues $0$ and $E$, and if
  $E\ell\leq\eta^{1/4}$ and one has that
  \begin{equation}
    \label{eq:42}
    \forall x\in[0,\ell],\quad r_u(x)\cdot r_v(x)\leq\frac{\eta}{\ell},
  \end{equation}
  then, there exists two points $x_+$ and $x_-$ in the lattice
  $\eta\Z$ satisfying
  \begin{equation}
    \label{eq:43}
    |\log\eta|/C<x_-<x_+<\ell-|\log\eta|/C\quad\text{and}\quad
    S< x_+-x_-<2S
  \end{equation}
  such that, if $H_-$ (resp. $H_+$) denotes the second order
  differential operator $H$ defined by~\eqref{eq:25} and Dirichlet
  boundary conditions on $[0,x_-]$ (resp. on $[x_+,\ell]$), then $H_-$
  and $H_+$ have an eigenvalue in the interval $[-E\ell\eta^{-1/4},
  E\ell\eta^{-1/4}]$.
\end{Pro}
\begin{proof}
  As $u$ and $v$ are normalized, one can pick $x_u$ (resp. $x_v$)
  s.t. $r_u(x_u)\geq\ell^{-1/2}$
  (resp. $r_v(x_v)\geq\ell^{-1/2}$). Thus, by~\eqref{eq:42}, one has
  $r_u(x_v)\leq\eta\ell^{-1/2}$ and $r_v(x_u)\leq\eta\ell^{-1/2}$.  To
  fix ideas, assume $x_u<x_v$. Note that, as $r_u$ satisfies
  equation~\eqref{eq:28}, one has $|\log\eta|/C\leq x_v-x_u$ (for some
  $C$ depending only on $\|q\|_\infty$). Hence, as $x\mapsto
  (r_u/r_v)(x)$ is continuous, there exists $x_u<x_0<x_v$ such that
  $r_u(x_0)=r_v(x_0)$. Define $x_\pm$ to be respectively the points in
  the lattice $\eta\Z$ closest to $x_0\pm S/2$. Then, there exists
  $C>0$ (depending only on $S$ and $\|q\|_\infty$) such that
  \begin{equation}
    \label{eq:44}
    \frac1C\leq (r_u/r_v)(x_\pm)\leq C\quad\text{ and
    }\quad|\log\eta|/C\leq\inf(x_v-x_+,x_--x_u).
  \end{equation}
  We will start with $H_-$ on the interval $[0,x_-]$; the case of
  $H_+$ on the interval $[x_+,\ell]$ is dealt with in the same
  way.\\
  Assume that $|\sin(\varphi_v(x_-))|\geq\sqrt[4]{\eta}$. Then, we
  pick $\D\lambda=\frac{u(x_-)}{v(x_-)}$ and set $w_-=u-\lambda v$.
  Thus, $w$ vanishes at the points $0$ and $x_-$ and, one computes
  \begin{equation*}
    \|H_-w_-\|_{L^2([0,x_-])}\leq E\lambda\leq
    \frac{E}{\sqrt[4]{\eta}}
  \end{equation*}
  and, using $r_u(x_u)\geq\ell^{-1/2}$ and~\eqref{eq:23} in
  Lemma~\ref{le:13}, for $\eta$ sufficiently small
  \begin{equation*}
    \begin{split}
      \|w\|^2_{L^2([0,x_-])}&=\int_0^{x_-}(u(x)-\lambda
      v(x))^2dx\\
      &\geq\int_0^{x_-}u^2(x)dx- 2\lambda\int_0^{x_-}r_u(x)r_v(x)dx
      \geq \ell^{-1}/C-\eta^{3/4}\ell^{-1}\geq \frac1{2C\ell}
    \end{split}
  \end{equation*}
  for $\eta$ sufficiently small. Hence, as $H_-$ is self-adjoint, we
  have proved the statement of Proposition~\ref{pro:1} if
  $|\sin(\varphi_v(x_-))|\geq\sqrt[4]{\eta}$. \\
  Assume now that $|\sin(\varphi_v(x_-))|\leq\sqrt[4]{\eta}$. Then,
  for $\eta$ sufficiently small, point (2) of Lemma~\ref{le:17} for
  $\varphi_v$ guarantees that, for some $x_0\in\eta\Z$ such that
  $x_--2\sqrt[8]{\eta}\leq x_0\leq x_--\sqrt[8]{\eta}$, one has
  $|\sin(\varphi_v(x_0))|\geq\sqrt[4]{\eta}$. So, we can
  do the computations done above replacing $x_-$ with $x_0$.\\
  To obtain the counterpart of this analysis for $H_+$ on
  $[x_+,\ell]$, we proceed as above except for the fact that we set
  $w_+=v-\lambda^{-1}u$ where $\lambda$ is chosen as before and
  estimate
  $\|(H_+-E)w_+\|_{L^2([x_+,\ell])}$.\\
  This completes the proof of Proposition~\ref{pro:1}.
\end{proof}
\noindent When we use Theorem~\ref{thr:7} to derive
Theorem~\ref{thr:1}, it will be of importance to have two points $x_-$
and $x_+$ that are well separated from each other. Sight changes in
the proof of Proposition~\ref{pro:1} yield the following result
\begin{Pro}
  \label{pro:2}
  There exists $\eta_0>0$ (depending only on $\|q\|_\infty$) such
  that, for $\eta\in(0,\eta_0)$ and $\ell$ sufficiently large, if $u$
  and $v$ are as section~\ref{sec:inverse-tunneling} and
  if~\eqref{eq:42} is satisfied, then, there exists a points
  $\overline{x}$ in the lattice $\eta\Z$ satisfying
  \begin{equation*}
    |\log\eta|/C<\overline{x}<\ell-|\log\eta|/C
  \end{equation*}
  such that, if $H_-$ (resp. $H_+$) denotes the second order
  differential operator $H$ defined by~\eqref{eq:25} and Dirichlet
  boundary conditions on $[0,\overline{x}]$ (resp. on
  $[\overline{x},\ell]$), then $H_-$ and $H_+$ have an eigenvalue in
  the interval $[-E\ell\eta^{-1/4},E\ell\eta^{-1/4}]$.
\end{Pro}
\subsubsection{Completing the proof of Theorem~\ref{thr:7}}
\label{sec:compl-proof-theor}
It suffices to pick $\eta$ so small that both Propositions~\ref{pro:3}
and~\ref{pro:1} hold. Recall that there is a change of notations
between Theorem~\ref{thr:7} and
Propositions~\ref{pro:3}~-~\ref{pro:1}. In Theorem~\ref{thr:7},
$E-\varepsilon$ (resp. $E+\varepsilon$) plays the role that $0$
(resp. $E$) plays in Propositions~\ref{pro:3} and~\ref{pro:1},
$2\varepsilon$ that of $E$ and $\varepsilon_0$ that of a power of
$\eta$ that is now fixed. \qed
\section{The proof of Theorems~\ref{thr:1}}
\label{sec:proof-theorem1}
\noindent The basic idea of the proof follows the basic idea
of~\cite{MR2775121} i.e. use localization to reduce the complexity of
the problem by reducing it to studying eigenvalues of $H_\omega$
restricted to cubes of size roughly $(\log L)^{1/\xi}$ for
$\xi\in(0,1)$.
\subsection{Reduction to localization cubes}
\label{sec:reduct-local-cubes}
Pick $J$ a compact interval where (Loc) is satisfied. Thus, we know
\begin{Le}[\cite{Ge-Kl:10}]
  \label{le:1}
  Under assumption (W) and (Loc), for any $\xi'\in(0,1)$ and
  $\xi''\in(0,\xi')$, for $L\geq1$ sufficiently large, with
  probability larger than $1-e^{-L^{\xi''}}$, if
  \begin{enumerate}
  \item $\varphi_{n,\omega}$ is a normalized eigenvector of
    $H_{\omega}(\Lambda_L)$ associated to $E_{n,\omega}\in J$,
  \item $x_n(\omega)\in \Lambda_L$ is a maximum of
    $x\mapsto\|\varphi_{n,\omega}\|^2_x=
    \int_{[x-1,x+1]\cap\Lambda_L}|\varphi_{n,\omega}(y)|^2dy$ in
    $\Lambda_L$,
  \end{enumerate}
  then, for $x\in\Lambda_L$, one has
  \begin{equation}
    \label{eq:68}
    \|\varphi_{n,\omega}\|_x\leq e^{2L^{\xi''}} e^{-|x-x_n(\omega)|^{\xi'}}.
  \end{equation}
\end{Le}
\noindent So, with good probability, all the eigenfunctions
essentially live in cubes of size of order $(\log L)^{1/\xi'}$ for any
$\xi'\in(0,1)$. Thus, they only see the configuration $\omega$ in such
cubes. To fix ideas, we define the center of localization of an
eigenfunction $\varphi$ as the left most maximum of
$x\mapsto\|\varphi\|_x$.
\begin{Rem}
  \label{rem:9}
  When (Loc) takes the form~\eqref{eq:81}, the estimate~\eqref{eq:68}
  can be replaced with $\D\|\varphi_{n,\omega}\|_x\leq e^{2L^{\xi'}}
  e^{-\xi|x-x_n(\omega)|}$.
\end{Rem}
\noindent We prove
\begin{Le}
  \label{le:3}
  Assume (W) and (Loc). Fix $J$ compact in $\overset{\circ}{I}$. Then,
  for any $\xi\in(0,1)$ and $\xi'\in(0,\xi)$, there exists
  $C=C_{\xi,\xi'}>0$ and $L_{\xi,\xi'}>0$ s.t., for $E\in J$, $L\geq
  L_{q,\xi}$ and $\varepsilon\in(0,1)$, one has
  \begin{multline}
    \label{eq:69}
    \sum_{k\geq2}\P\left(\tr[\car_{[E-\varepsilon,E+\varepsilon]}
      (H_\omega(\Lambda_L))]\geq k\right)\\\leq
    e^{-s\ell^{\xi'}/9}+\frac{L^2}{\ell}\P_{2,9\ell,\ell}
    \left(\varepsilon\right) \\ + \left(\frac{L}{\ell}\right)^2
    \left(\P_{1,3\ell/2,4\ell/3}(\varepsilon)+e^{-\ell^{\xi'}/8}\right)^2
    e^{L\,\P_{1,3\ell/2,4\ell/3}(\varepsilon)/\ell}
  \end{multline}
  where $\ell=(\log L)^{1/\xi}$ and, for $j\geq1$, one has set
  \begin{equation}
    \label{eq:13}
    \P_{j,\ell,\ell'}(\varepsilon):=\sup_{\gamma\in\ell'\Z\cap[0,L]}
    \P\left(\tr[\car_{[E-\varepsilon,E+\varepsilon]}
      (H_\omega(\Lambda_\ell(\gamma)))]\geq j\right).
  \end{equation}
\end{Le}
\begin{Rem}
  \label{rem:9}
  When (Loc) takes the form~\eqref{eq:81}, in Lemma~\ref{le:3}, one
  can pick $\ell=K\log L$ with $K$ sufficiently large.
\end{Rem}
\begin{proof}[Proof of Lemma~\ref{le:3}]
  Pick $E\in J$. First, by standard bounds on the eigenvalue counting
  function of $-\Delta$, there exists $C>0$ depending only on $J$ such
  that, for $\varepsilon\in(0,1)$, one has
  \begin{equation}
    \label{eq:9}
    \tr[\car_{[E-\varepsilon,E+\varepsilon]}(H_\omega(\Lambda_L))]\leq CL.
  \end{equation}
  Pick $\xi'\in(\xi,1)$ and $\xi''\in(0,\xi)$. Let
  $\mathcal{Z}_{\xi',\xi''}$ be the set of configurations $\omega$
  defined by Lemma~\ref{le:1} for the exponents $\xi'$ and $\xi''$. It
  has probability at least $1-e^{-L^{\xi''}}$. Thus, by~\eqref{eq:9},
  we estimate, for $L$ sufficiently large,
  \begin{equation}
    \label{eq:70}
    \sum_{k\geq2}\P\left(\left\{\omega\not\in\mathcal{Z}_{\xi',\xi''};\
        \tr[\car_{[E-\varepsilon,E+\varepsilon]}
        (H_\omega(\Lambda_L))]\geq k\right\}\right)
    \leq C L e^{-L^{\xi''}}\leq e^{-\ell^{\xi'}}
  \end{equation}
  as $\ell=(\log L)^{1/\xi}$.\\
  Let us now estimate $\P(\left\{\omega\in\mathcal{Z}_{\xi',\xi''};\
    \tr[\car_{[E-\varepsilon,E+\varepsilon]}(H_\omega(\Lambda_L))]\geq
    k\right\})$.\\
  For $\omega\in\mathcal{Z}_{\xi',\xi''}$, by Lemma~\ref{le:1}, for
  each $\varphi$ eigenfunction of $H_\omega(\Lambda_L)$ associated to
  an eigenvalue $E\in J$, we define the center of localization
  associated to $\varphi$ as in the remarks following
  Lemma~\ref{le:1}. We consider the events
  $\Omega_{\xi',\xi''}^b:=\mathcal{Z}_{\xi',\xi''}
  \setminus\Omega_{\xi',\xi''}^g$ and
  \begin{equation*}
    \Omega_{\xi',\xi''}^g:=\left\{\omega\in\mathcal{Z}_{\xi',\xi''};\
      \begin{aligned}
        \text{no two centers of
          localization of eigenfunctions}\\
        \text{associated to eigenvalues
          in }[E-\varepsilon,E+\varepsilon]\\
        \text{are at a distance less than }4\ell\text{ from each
          other}
      \end{aligned}
    \right\}.
  \end{equation*}
  Note that, for $\omega\in\Omega_{\xi',\xi''}^g$,
  $H_\omega(\Lambda_L)$ has at most $[L/(4\ell)]+1$ eigenvalues in
  $[E-\varepsilon,E+\varepsilon]$;
  here, $[\cdot]$ denotes the integer part of $\cdot$.\\
  We prove
  \begin{Le}
    \label{le:2}
    Fix $0<\xi''<\xi<\xi'<1$. Then, there exists
    $L_{\xi,\xi',\xi''}>0$ such that, for $\ell=(\log L)^{1/\xi}$, for
    $L\geq L_{\xi,\xi',\xi''}$ and $k\geq2$, one has
    \begin{equation}
      \label{eq:10}
      \P\left(\left\{\omega\in\Omega^b_{\xi',\xi''};\
          \tr[\car_{[E-\varepsilon,E+\varepsilon]}(H_\omega(\Lambda_L))]
          \geq k\right\}\right)\leq \frac{L}{\ell}\P_{2,9\ell,\ell}
      \left(\varepsilon\right)+e^{-s\ell^{\xi'}/9}
    \end{equation}
    and, for $k\leq[L/(4\ell)]+1$,
    \begin{multline}
      \label{eq:12}
      \P\left(\left\{\omega\in\Omega^g_{\xi',\xi''};\
          \tr[\car_{[E-\varepsilon,E+\varepsilon]}(H_\omega(\Lambda_L))]\geq
          k\right\}\right)\\\leq \binom{[L/\ell]}{k}
      \left(\P_{1,3\ell/2,4\ell/3}(\varepsilon)
        +e^{-\ell^{\xi'}/8}\right)^k
    \end{multline}
    where $\P_{j,\ell,\ell'}(\varepsilon)$ is defined
    in~\eqref{eq:13}.
  \end{Le}
  \noindent We postpone the proof of Lemma~\ref{le:2} to complete that
  of Lemma~\ref{le:3}. We pick $q\geq1$ and sum~\eqref{eq:10}
  and~\eqref{eq:12} for $k\geq2$ to get, for some $C>0$
  \begin{equation*}
    \begin{split}
      &\frac1C\sum_{k\geq2}\P\left(\tr[\car_{[E-\varepsilon,E+\varepsilon]}
        (H_\omega(\Lambda_L))]\geq k\right)\\&\leq
      \left(\frac{L}{\ell}\right)^2
      \left(\P_{1,3\ell/2,4\ell/3}(\varepsilon)+ e^{-(|\log
          L|)^{\xi'/\xi}/8}\right)^2 \left(
        1+\P_{1,3\ell/2,4\ell/3}(\varepsilon)+e^{-(|\log
          L|)^{\xi'/\xi}/8} \right)^{L/\ell} \\ &\hskip8cm+
      e^{-s\ell^{\xi'}/9}
      +\frac{L^2}{\ell}\P_{2,9\ell,\ell}\left(\varepsilon\right)\\
      &\leq C\left( \left(\frac{L}{\ell}\right)^2
        \left(\P_{1,3\ell/2,4\ell/3}(\varepsilon)+e^{-(|\log
            L|)^{\xi'/\xi}/8}\right)^2
        e^{L\,\P_{1,3\ell/2,4\ell/3}(\varepsilon)/\ell}\right.\\
      &\hskip8cm \left. +e^{-s\ell^{\xi'}/9}
        +\frac{L^2}{\ell}\P_{2,9\ell,\ell}\left(\varepsilon\right)
      \right).
    \end{split}
  \end{equation*}
  Here, we have used the following bound, for $(x,y)\in(\R^+)^2$ and
  $m\leq n$ integers,
  \begin{equation}
    \label{eq:16}
    \sum_{k=m}^n\binom{n}{k}x^ky^{n-k}\leq\binom{n}{m}x^m(x+y)^{n-m}.
  \end{equation}
  This completes the proof of Lemma~\ref{le:3}.
\end{proof}
\begin{proof}[Proof of Lemma~\ref{le:2}]
  We will use
  \begin{Le}
    \label{le:18}
    For $0<\xi''<\xi<\xi'<1$, there exists $L_{\xi,\xi',\xi''}>0$ such
    that for $\ell=(\log L)^{1/\xi}$ and $L\geq L_{\xi,\xi',\xi''}$
    and $\omega\in\mathcal{Z}_{\xi',\xi''}$, for any
    $\gamma\in\Lambda_L$, if $H_\omega(\Lambda_L)$ has $k$ eigenvalues
    in $[E-\varepsilon,E+\varepsilon]$ with localization center in
    $\Lambda_{4\ell/3}(\gamma)$, then
    $H_\omega(\Lambda_{3\ell/2}(\gamma))$ has $k$ eigenvalues in
    $\left[E-\varepsilon-e^{-\ell^{\xi'}/8},E+\varepsilon
      -e^{-\ell^{\xi'}/8}\right]$.
  \end{Le}
  \noindent We postpone the proof of Lemma~\ref{le:18} to complete
  that of Lemma~\ref{le:2}.  Pick $k\geq2$. We first estimate
  $\D\P\left(\left\{\omega\in\Omega^b_{\xi',\xi''};\
      \tr[\car_{[E-\varepsilon,E+\varepsilon]}(H_\omega(\Lambda_L))]\geq
      k\right\}\right)$. Clearly, one has
  \begin{multline*}
    \P\left(\left\{\omega\in\Omega^b_{\xi',\xi''};\
        \tr[\car_{[E-\varepsilon,E+\varepsilon]}(H_\omega(\Lambda_L))]\geq
        k\right\}\right)\\\leq
    \P\left(\left\{\omega\in\Omega^b_{\xi',\xi''};\
        \tr[\car_{[E-\varepsilon,E+\varepsilon]}(H_\omega(\Lambda_L))]\geq
        2\right\}\right).
  \end{multline*}
  Thus, we take $k=2$.\\
  By the definition of $\Omega^b_{\xi',\xi''}$ and Lemma~\ref{le:18},
  one clearly has
  \begin{equation*}
    \begin{split}
      &\P\left(\left\{\omega\in\Omega^b_{\xi',\xi''};\
          \tr[\car_{[E-\varepsilon,E+\varepsilon]}(H_\omega(\Lambda_L))]\geq
          2\right\}\right)\\ &\leq
      \P\left(\left\{\exists\gamma\in\ell\Z\cap[0,L];\
          \tr[\car_{[E-\varepsilon-e^{-\ell^{\xi'}/8},E+\varepsilon
            +e^{-\ell^{\xi'}/8}]}
          (H_\omega(\Lambda_{9\ell}(\gamma)))]\geq 2\right\}\right) \\
      &\leq \sum_{\gamma\in\ell\Z\cap[0,L]}\P\left(\left\{
          \tr[\car_{[E-\varepsilon-e^{-\ell^{\xi'}/8},
            E+\varepsilon+e^{-\ell^{\xi'}/8}]}
          (H_\omega(\Lambda_{9\ell}(\gamma)))]\geq 2\right\}\right) \\
      &\leq \frac{L}{\ell}\P_{2,9\ell,\ell}\left(\varepsilon
        +e^{-\ell^{\xi'}/8}\right)\leq
      \frac{L}{\ell}\P_{2,9\ell,\ell}\left(\varepsilon\right)+
      e^{-s\ell^{\xi'}/9}
    \end{split}
  \end{equation*}
  for $L$ sufficiently large as $\ell=(\log L)^{1/\xi}$ and
  $\xi'>\xi$; in the last step, we have used the Wegner estimate
  (W). This completes the proof of~\eqref{eq:10}.\\
  Let us now estimate $\P\left(\left\{\omega\in\Omega^g_{\xi',\xi''};\
      \tr[\car_{[E-\varepsilon,E+\varepsilon]}(H_\omega(\Lambda_L))]\geq
      k\right\}\right)$. We cover the cube $\Lambda_L$ by cubes
  $(\Lambda_{4\ell/3}(\gamma))_{\gamma\in\Gamma}$
  i.e. $\Lambda_L=\cup_{\gamma\in\Gamma}\Lambda_{4\ell/3}(\gamma)$ in
  such a way that $[3L/(4\ell)]\leq\#\Gamma\leq[L/\ell]$.\\
  Assume now that $\omega\in\Omega^g_{\xi',\xi''}$ is such that $\tr[
  \car_{[E-\varepsilon,E+\varepsilon]}(H_\omega(\Lambda_L))]\geq k$.
  Thus, the localization centers for any two eigenfunctions being at
  least $4\ell$ away from each other, by Lemma~\ref{le:18}, we can
  find $k$ points in $\Gamma$, say $(\gamma_j)_{1\leq j\leq k}$ such
  that
  \begin{itemize}
  \item for $1\leq j\leq k$, $H_\omega(\Lambda_{3\ell/2}(\gamma_j))$
    has exactly one eigenvalue in the interval
    $\left[E-\varepsilon-e^{-\ell^{\xi'}/8},
      E+\varepsilon+e^{-\ell^{\xi'}/8}\right]$;
  \item for $1\leq j<j'\leq k$, one has
    dist$(\Lambda_{3\ell/2}(\gamma_j),
    \Lambda_{3\ell/2}(\gamma_{j'}))>\ell/2$.
  \end{itemize}
  Hence, by (IAD), for $\ell$ sufficiently large, the operators
  $(H_\omega(\Lambda_{3\ell/2}(\gamma_j)))_{1\leq j\leq k}$ are
  stochastically independent. Hence, we have the bound
  \begin{multline*}
    \P\left(\left\{\omega\in\Omega^g_{\xi',\xi''};\
        \tr[\car_{[E-\varepsilon,E+\varepsilon]}(H_\omega(\Lambda_L))]
        \geq k\right\}\right)\\\leq \binom{\#\Gamma}{k}
    \left(\P_{1,3\ell/2,4\ell/3}(\varepsilon) +e^{-(|\log
        L|)^{\xi'/\xi}/8}\right)^k .
  \end{multline*}
  As $[3L/(4\ell)]\leq\#\Gamma\leq[L/\ell]$ and $k\leq [L/(4\ell)]$,
  this completes the proof of~\eqref{eq:12} and, thus, of
  Lemma~\ref{le:2}.
\end{proof}
\begin{proof}[Proof of Lemma~\ref{le:18}]
  Analogous results can be found in~\cite{MR2775121,Ge-Kl:10}.\\
  If $\varphi$ is an eigenfunction of $H_\omega(\Lambda_L)$ associated
  to $e$ an eigenvalue in $[E-\varepsilon,E+\varepsilon]$ that has
  localization center in $\Lambda_{4\ell/3}(\gamma)$, then, by
  \eqref{eq:68} in Lemma~\ref{le:1}, we have that, for $\chi$ a smooth
  cut-off that is $1$ on $\Lambda_{10\ell/9}(\gamma)$ and vanishing
  outside $\Lambda_{3\ell/2}(\gamma)$, one has, for $L$ sufficiently
  large,
  \begin{equation*}
    \left\|H_\omega(\Lambda_{3\ell/2}(\gamma))-e)(\chi\varphi)\right\|
    \leq e^{2\ell^{\xi''}}e^{-(\ell/6)^{\xi'}}\leq 
    e^{-\ell^{\xi'}/8}.
  \end{equation*}
  Recall that $\xi'/\xi>1$. On the other hand, if one has $k$ such
  eigenvalues, say, $(\varphi_j)_{1\leq j\leq k}$, then $k\leq C L$
  and one computes the Gram matrix in the same way
  \begin{equation*}
    \begin{split}
      ((\langle \chi\varphi_j,\chi\varphi_{j'}\rangle))_{1\leq
        j,j'\leq k}&=((\langle\varphi_j,\varphi_{j'}\rangle))_{1\leq
        j,j'\leq k}+O\left(k^2 e^{-\ell^{\xi'}/8}\right)\\
      &=\text{Id}_k+O\left(L^2 e^{-\ell^{\xi'}/8}\right).
    \end{split}
  \end{equation*}
  as $k$ is bounded by $CL$. This completes the proof of
  Lemma~\ref{le:18}.
\end{proof}
\subsection{The proof of Theorem~\ref{thr:1}}
\label{sec:proof-theorem4}
We use Lemma~\ref{le:3}. Recall that $\ell=(\log
L)^{1/\xi}$. In~\eqref{eq:69}, to estimate
$\P_{1,3\ell/2,4\ell/3}(2\varepsilon)$, we use the Wegner type
estimate (W) and obtain
\begin{equation}
  \label{eq:17}
  \P_{1,3\ell/2,4\ell/3}(\varepsilon)\leq C \varepsilon^s(\log
  L)^{\rho/\xi}.
\end{equation}
To estimate $\P_{2,9\ell,\ell}(2\varepsilon)$, we use
Theorem~\ref{thr:7} and the Wegner type estimate (W). The point
$(x_\pm)$ are not known but we know that they belong to the lattice
segment $\varepsilon_0\Z\cap[0,\ell]$ (independent of the potential
$q_\omega$) so there are at most $(\ell/\varepsilon_0)^2$ possible
pairs of points. We choose the constant $S>R_0$ defined by (IAD);
hence, as the points $x_+-x_-\geq S$, the operators
$H_-:=H^D_{\omega|[0,x_-]}$ and $H_+:=H^D_{\omega|[x_+,\ell]}$ are
stochastically independent. Thus, applying the Wegner type estimate
(W) for the operators $H_\pm$ and summing over the pairs of points in
$\varepsilon_0\Z\cap[0,\ell]$ yields
\begin{equation}
  \label{eq:11}
  \P_{2,9\ell,\ell}(2\varepsilon)\leq 
  C(\log L)^{2/\xi}\left(\varepsilon(\log L)^{4/\xi}\right)^{2s}.
\end{equation}
Plugging this and~\eqref{eq:17} into~\eqref{eq:69} yields~\eqref{eq:8}
with
\begin{equation*}
  \eta:=\xi'/\xi,\quad
  \beta:=\max(1+4s,\rho)/\xi'=\frac{\max(1+4s,\rho)}{\eta\xi}
  \quad\text{and}\quad
  \rho':=\rho/\xi'=\frac\rho{\eta\xi}.
\end{equation*}
As $\xi<\xi'<1$ can be chosen arbitrary, this completes the proof of
Theorem~\ref{thr:1}. \qed
\section{Proofs of the universal estimates}
\label{sec:universal-estimates}
\noindent We now prove Theorems~\ref{thr:3} and~\ref{thr:6}. By a
shift in energy, it suffices to prove the results for $E=0$ and see
that the constants only depend on $\|q\|_\infty$. From now on, we
assume the energy interval under consideration is centered at $E=0$.
\begin{proof}[Proof of Theorem~\ref{thr:3}]
  Pick $\varepsilon\in(0,1)$. Assume $H$ has at least two eigenvalues,
  say, $E$ and $\tilde E$ in $[-\varepsilon,\varepsilon]$. By shifting
  the potential by a constant less than $1$, without loss of
  generality, we may assume that $\tilde E=0$ and $E>0$. Let $v$ and
  $w$ be the fundamental solutions to the equation $-u''+qu=0$
  (i.e. $v(0)=1=w'(0)$ and $v'(0)=0=w(0)$) and let $S_0(y,x)$ be the
  resolvent matrix associated to $(v,w)$ i.e.
  \begin{equation*}
    S_0(y,x)=
    \begin{pmatrix} v(y)& w(y)\\v'(y) &w'(y) \end{pmatrix}
    \begin{pmatrix} w'(x)& -w(x)\\-v'(x) &v(x) \end{pmatrix}.
  \end{equation*}
  Clearly $S_0$ solves
  \begin{equation*}
    \frac{d}{dy}S_0(y,x)=
    \begin{pmatrix} 0& 1\\ q(y)&0\end{pmatrix}S_0(y,x), \quad S_0(x,x)=
    \begin{pmatrix}1&0\\0&1\end{pmatrix}.
  \end{equation*}
  Obviously, as $q$ is bounded, for some $C$ depending only on
  $\|q\|_\infty$, one has
  \begin{equation}
    \label{eq:2}
    \|S_0(y,x)\|\leq e^{C|y-x|}.
  \end{equation}
  Let $u$ be a $L^2([0,\ell])$-normalized solution to $Hu=Eu$. Hence,
  we have
  \begin{equation}
    \label{eq:63}
    \begin{pmatrix} u(x)\\u'(x)\end{pmatrix}=S_0(x,0)
    \begin{pmatrix}
      u(0)\\u'(0)\end{pmatrix}+\int_0^xS_0(y,0)B(y)dy
  \end{equation}
  where
  \begin{equation*}
    B(y)=E\begin{pmatrix} 0\\u(x)\end{pmatrix}.
  \end{equation*}
  The eigenfunction, say, $u_0$, associated to $H$ and $0$ can be
  written as
  \begin{equation*}
    \begin{pmatrix} u_0(x)\\u_0'(x)\end{pmatrix}=S_0(x,0)
    \begin{pmatrix} u_0(0)\\u_0'(0)\end{pmatrix}.
  \end{equation*}
  As $u$ and $u_0$ satisfy the same boundary conditions,
  using~\eqref{eq:63},~\eqref{eq:2} and the normalization of $u$, we
  get that, for some $\lambda>0$, one has
  \begin{equation}
    \label{eq:3}
    \left\|\begin{pmatrix} u\\u'\end{pmatrix}-
      \lambda\begin{pmatrix} u_0\\u_0'\end{pmatrix}\right\|_\infty\leq
    C  \varepsilon e^{C\ell}.
  \end{equation}
  If $\varepsilon\in(0,1)$ such that $|\log\varepsilon|\geq K\ell$
  where $K$ is taken such that, for $\ell\geq1$, one has
  $Ce^{(C-K)\ell}\ell<1$. By~\eqref{eq:3}, as, on $[0,\ell]$, $u$ and
  $u_0$ are normalize and orthogonal to each other, we get
  $\lambda^2+1<1$ which is absurd. This completes the proof of
  Lemma~\ref{thr:3}.
\end{proof}
\begin{proof}[Proof of Theorem~\ref{thr:6}]
  Assume $H$ has $N+1$ eigenvalues in $[-\varepsilon,\varepsilon]$.
  As $q$ is bounded, standard comparison with the Laplace operator
  $H_0=-d^2/dx^2$ implies that $N\leq C\ell$ for some $C>0$ depending
  only $\|q\|_\infty$.\\
  As in the proof of Theorem~\ref{thr:3}, we may assume that the
  smallest one of them be $0$, thus, that the other be positive. Let
  $(u_j)_{0\leq j\leq N}$ be the associated normalized eigenfunctions,
  $u_0$ being the one associated to the eigenvalue
  $0$.\\
  Fix $1\leq\tilde\ell<\ell$ to be chosen later. Partition the
  interval $[0,\ell]$ into $A$ intervals of length approximately
  $\tilde\ell$ i.e.  $\D [0,\ell]=\cup_{1\leq \alpha\leq A}I_\alpha$
  where $I_\alpha=[x_\alpha,x_{\alpha+1}]$ and
  $x_{\alpha+1}-x_\alpha\asymp\tilde\ell$; hence, $A\asymp
  \ell/\tilde\ell$.\\
  As in Lemma~\ref{thr:3}, let $(v,w)$ be the fundamental solutions to
  $-u''+qu=0$. Formula~\eqref{eq:63} and~\eqref{eq:2} show that there
  exists constants $\D((\lambda_j^\alpha))_{\substack{1\leq j\leq
      N\\1\leq \alpha\leq A}}$ and
  $\D((\beta_j^\alpha))_{\substack{1\leq j\leq N\\1\leq \alpha\leq
      A}}$ such that, for $0\leq j\leq N$ and $1\leq \alpha\leq A$, we
  have
  \begin{equation}
    \label{eq:4}
    \sup_{x\in I_\alpha}\left|\begin{pmatrix} u_j(x)\\u_j'(x)\end{pmatrix}-
      \lambda^\alpha_j\begin{pmatrix} v(x)\\v'(x)\end{pmatrix}
      +\beta^\alpha_j\begin{pmatrix} w(x)\\w'(x)\end{pmatrix}\right|\leq
    C\varepsilon e^{C\tilde\ell}.
  \end{equation}
  Let $\langle\cdot,\cdot\rangle$ denote the standard scalar product
  on $L^2([0,\ell])$ and $\langle\cdot,\cdot\rangle_\alpha$ that on
  $L^2(I_\alpha)$. One has
  \begin{equation}
    \label{eq:5}
    \text{Id}_{N+1}=\left(\left(\langle u_i,u_j\rangle\right)
    \right)_{\substack{0\leq i\leq N\\0\leq j\leq N}}
    =\sum_{\alpha=1}^A    
    \left(\left(\langle u_i,u_j\rangle_\alpha\right)
    \right)_{\substack{0\leq i\leq N\\0\leq j\leq N}}.
  \end{equation}
  Using~\eqref{eq:4}, we compute
  \begin{equation}
    \label{eq:6}
    M_\alpha:=\left(\left(\langle u_i,u_j\rangle_\alpha\right)
    \right)_{\substack{0\leq i\leq N\\0\leq j\leq
        N}}=\sum_{n=1}^4 M_{\alpha,n}+S_\alpha      
  \end{equation}
  where
  \begin{gather}
    \label{eq:65}
    M_{\alpha,1}=\langle v,v\rangle_\alpha\left(\left(\lambda_i^\alpha
        \lambda_j^\alpha\right) \right)_{\substack{0\leq i\leq
        N\\0\leq j\leq N}},\quad M_{\alpha,2}=\langle
    v,w\rangle_\alpha\left(\left(\lambda_i^\alpha
        \beta_j^\alpha\right) \right)_{\substack{0\leq i\leq N\\0\leq
        j\leq N}},\\\label{eq:72} M_{\alpha,3}= \langle
    w,v\rangle_\alpha\left(\left(\beta_i^\alpha
        \lambda_j^\alpha\right) \right)_{\substack{0\leq i\leq
        N\\0\leq j\leq N}},\quad M_{\alpha,4}= \langle
    w,w\rangle_\alpha\left(\left(\beta_i^\alpha \beta_j^\alpha\right)
    \right)_{\substack{0\leq i\leq N\\0\leq j\leq N}}, \\\label{eq:71}
    \|S_\alpha\|\leq C\varepsilon N e^{C\tilde\ell}\tilde\ell\leq
    C\varepsilon\, \ell\,\tilde\ell\, e^{C\tilde\ell}.
  \end{gather}
  Pick $\tilde\ell=|\log\varepsilon|/K$ for some $K$ sufficiently
  large; as $0<\varepsilon\leq\ell^{-\nu}$ with $\nu>2$, for $\ell$
  sufficiently large, by~\eqref{eq:71}, one has
  \begin{equation*}
    \sum_{\alpha=1}^A\|S_\alpha\|\leq C\varepsilon \ell^2
    e^{C\tilde\ell}\leq C\ell^{2-\nu(1-C/K)}\leq1/2.
  \end{equation*}
  By~\eqref{eq:65} and~\eqref{eq:72}, the matrices
  $(M_{\alpha,n})_{\alpha,n}$ are all of rank at most $1$. Hence,
  equation~\eqref{eq:5} implies that $4A\geq N+1$ which yields
  $N+1\leq C\ell/|\log\varepsilon|$ for some $C>0$. This completes the
  proof of Theorem~\ref{thr:6}.
\end{proof}
\noindent One can wonder whether the bounds given in
Theorems~\ref{thr:3} and~\ref{thr:6} are optimal. Examples build using
semi-classical ideas show that the orders of magnitudes are. The
precise values of the constants depend on the details of the potential
$q$.
\section{Localization for the models $H_\omega^A$ and $H_\omega^D$}
\label{sec:localization-models}
In the present section, we establish that the models $H_\omega^A$ and
$H_\omega^D$ satisfy (Loc) as claimed in the introduction.
\subsection{Localization for the model $H_\omega^A$}
\label{sec:local-model-h_om}
In the present section, we show how to extend the results
of~\cite{0912.3568} to our assumptions.\\
Let
\begin{equation*}
  \tilde H_\omega=-\frac{d^2}{dx^2}+\tilde
  W(\cdot)+\sum_{n\in\Z}\tilde\omega_n \tilde V(\cdot-n)
\end{equation*}
where
\begin{itemize}
\item $(\tilde\omega_n)_{n\in\Z}$ and $\tilde V$ satisfy the
  assumptions that $(\omega_n)_{n\in\Z}$ and $V$ satisfy for
  $H^A_\omega$ in the introduction, section~\ref{intro},
\item $\tilde V$ has its support in $(-1/2,1/2)$,
\item $\tilde W$ is uniformly continuous on $\R$.
\end{itemize}
Then, the main result of~\cite{0912.3568} can be rephrased in the
following way: $H_\omega$ satisfies (Loc) (see~\eqref{eq:84}) for any
compact interval $I$ (see Lemma 2.1 and Proposition 2.2
in~\cite{0912.3568}).\\
Consider now $H^A_\omega$ as defined in section~\ref{intro}. Let
$n_0\in\N$ be such that supp$V\subset(-n_0/2,n_0/2)$. Doing the change
of variable $x=n_0 y$, we can rewrite
\begin{equation}
  \label{eq:21}
  H^A_\omega=n_0^{-2}\left(
    -\frac{d^2}{dy^2}+\tilde W(\cdot)+\sum_{n\in\Z}\tilde\omega_n
    \tilde V(\cdot-n)
  \right)
\end{equation}
where
\begin{itemize}
\item $\tilde V(\cdot)=n_0^2 V(n_0\,\cdot)$, thus, $\tilde V$ has its
  support in $(-1/2,1/2)$,
\item $\tilde\omega_n=\omega_{n\cdot n_0}$ for $n\in\Z$,
\item $\D\tilde W(\cdot)=n_0^2\sum_{n\in\Z\setminus n_0\Z}\omega_n
  V(n_0\,\cdot-n)$, thus, $\tilde W$ is uniformly continuous on $\R$
  for any realization $(\omega_n)_{n\in\Z\setminus n_0\Z}$ (as the
  random variables are bounded).
\end{itemize}
So, for any realization $(\omega_n)_{n\in\Z\setminus n_0\Z}$, we know
that $H^A_\omega$ satisfies assumption (Loc) on any compact interval
$I$ when the expectation is taken with respect to the random variables
$(\omega_n)_{n\in n_0\Z}$. A priori, the constant in the right hand
side of~\eqref{eq:84} may depend on the realization
$(\omega_n)_{n\in\Z\setminus n_0\Z}$. The proof of Theorem 1
in~\cite{0912.3568} shows that this is not the case. More precisely,
as $\tilde W$ stays uniformly bounded independently of the realization
$(\omega_n)_{n\in\Z\setminus n_0\Z}$, the estimates of the operator
$T_1$ and its continuity with respect to the potential $\tilde W$
($W_0$ in~\cite{0912.3568}) yield that the right hand side
of~\eqref{eq:84} is bounded uniformly in $(\omega_n)_{n\in\Z\setminus
  n_0\Z}$. Thus, $H^A_\omega$ satisfies (Loc) on any compact interval
$I$.
\subsection{Localization for the model $H_\omega^D$}
\label{sec:local-model-h_d}
The purpose of this section is to prove that, in the setting of the
introduction, there exists $\tilde E^D>\inf\Sigma^D$ such that
assumption~\eqref{eq:84} is satisfied in $(\inf\Sigma^D,\tilde E^D]$
for $H_\omega^D$. Actually we will prove this under assumptions weaker
than those made in the introduction. \\
Consider the random displacement model~\eqref{eq:61} where
\begin{itemize}
\item $V:\ \R\to\R$ is a smooth, even function that is compactly
  supported in $(-r_0,r_0)$ for some $0<r_0<1/2$;
\item $(\omega_n)_{n\in\Z}$ are bounded i.i.d random variables, the
  common distribution of which admits a density supported in
  $[-r,r]\subset [-1/2+r_0,1/2-r_0]$, that is continuously
  differentiable in $[-r,r]$ and which support contains $\{-r,r\}$.
\end{itemize}
For $a\in [-r,r]$, consider $H_1(a) = -\Delta + q(x-a)$ on
$L^2(-1/2,1/2)$ with Neumann boundary condition and let $E_0(a) = \inf
\sigma(H_1(a))$ be the lowest eigenvalue of $H_1(a)$. Note that, by
symmetry of $q$, $E_0(a)$ is even.\\
We prove
\begin{Th}
  \label{thr:2}
  Assume that $E_0(a)$ does not vanish identically for $a\in [-r,r]$.\\
  Then, there exists $\delta>0$ such that $H_{\omega}$ almost surely
  has pure point spectrum in $I=[E_0, E_0+\delta]$ with exponentially
  decaying eigenfunctions. Moreover, $H_{\omega}$ is dynamically
  localized in $I$, in the sense that for every
  $\zeta<1$,~\eqref{eq:84} holds.
\end{Th}
\noindent In~\cite{MR2503172}, it is proved that if $V$ has a fixed
sign, then $E_0(a)$ does not vanish identically for $a\in
[-r,r]$. Thus, under our assumptions in the introduction, we obtain
that assumption (Loc) holds in some neighborhood of the bottom of the
spectrum of $H_\omega^D$.\\
In~\cite{1007.2483}, Theorem~\ref{thr:2} was proved when
$d\geq2$. Here, we are going to extend the ideas used to prove it to
the one-dimensional case. \\
The proof of Theorem~\ref{thr:2} follows a well known strategy: to
prove localization in some energy region $I$, one only needs to prove
that, in $I$, the operator satisfies a Wegner estimate and the
resolvent of its restriction to a finite cube satisfies a smallness
estimate with a good probability (see e.g.~\cite[Theorem
5.4]{MR2509111}). This strategy is the one followed
in~\cite{1007.2483} that we also follow below.\\
For any $s\in(0,1)$ and $\rho=1$, the Wegner estimate (W) for our
model was proved in~\cite[Theorem 4.1]{1007.2483} under no restriction
on the dimension. In dimension 1, the same analysis can be improved to
give
\begin{Th}
  \label{thm:wegner}
  There exists $\delta>0$ and $C>0$ such that, for any $L>1$ and any
  interval $I\subset [\inf\Sigma^D,\inf\Sigma^D+\delta]$, one has
  \begin{equation}
    \label{eq:wegnerest}
    \E(\tr\chi_I(H^D_{\omega,L})) \le C |I| L.
  \end{equation}
  Thus, in $[\inf\Sigma^D,\inf\Sigma^D+\delta]$, the integrated
  density of states $E\mapsto N^D(E)$ is Lipschitz continuous.
\end{Th}
\noindent To obtain~\eqref{thm:wegner}, it suffices to follow the
proof of~\cite[Theorem 4.1]{1007.2483} and in ~\cite[(53)]{1007.2483}
to use the boundedness of the spectral shift function
$E\mapsto\xi(E;-\Delta+V,-\Delta+V+V_0)$ in dimension $1$ when $V$ is
bounded, and $V_0$ is bounded and of compact support (see~\cite[Remark
3.1]{MR1935272}).\\
Recall that $N(E)=N^D(E)$ denotes the integrated density of states of
$H_\omega^D$ (see~\eqref{eq:19}). The proof of the ``smallness'' of
the resolvent usually relies on a so-called ``Lifshitz tail'' estimate
for $N(E)$. Such an estimate says roughly that, at the bottom of the
spectrum (resp. at a so called fluctuational edge of the almost sure
spectrum (see e.g.~\cite{MR94h:47068})), the function $E\mapsto N(E)$
vanishes very
quickly (resp. is very flat).\\
In dimension $1$, in~\cite[Theorem 4.1]{MR2503172}, it was proved that
such a quick vanishing of $N$ fails for displacement model
$H_\omega^D$ when the random variables $(\omega_n)_{n\in\Z}$ have a
Bernoulli distribution supported in $\{-r,r\}$. It was also
conjectured that, when this is not the case, the integrated density of
states should be infinitely flat at $\inf\Sigma^D$. This is not the
case. Indeed, we prove that, if we assume $V$ to be as above and that
\begin{itemize}
\item $(\omega_n)_{n\in\Z}$ are i.i.d random variables supported in
  $[-r,r]\subset [-1/2+r_0,1/2-r_0]$ which support contains
  $\{-r,r\}$.
\end{itemize}
then one has
\begin{Th}
  \label{thr:12}
  In the above setting assume that $\P(\omega_0=r)\P(\omega_0=-r)>0$.\\
  Then, there exists $n\geq0$ such that one has
  \begin{equation}
    \label{eq:82}
    \lim_{\substack{E\to\inf\Sigma^D\\E>\inf\Sigma^D}}
    \frac{N(E)}{(E-\inf\Sigma^D)^n}\to+\infty.
  \end{equation}
\end{Th}
\noindent Under the same conditions on $V$ and $(\omega_n)_{n\in\Z}$,
we also prove
\begin{Th}
  \label{thr:8}
  In the above setting assume that $\P(\omega_0=r)+\P(\omega_0=-r)=0$.\\
  Then, for any $n\geq0$, one has
  \begin{equation}
    \label{eq:62}
    \lim_{\substack{E\to\inf\Sigma^D\\E>\inf\Sigma^D}}
    \frac{N(E)}{(E-\inf\Sigma^D)^n}=0.
  \end{equation}
\end{Th}
\noindent Theorem~\ref{thr:8} is not optimal: for it to be
optimal,~\eqref{eq:62} should hold under the weaker assumption
$\P(\omega_0=r)\P(\omega_0=-r)=0$.\\
Let us now complete the proof of Theorem~\ref{thr:2} using
Theorem~\ref{thr:8}. Clearly, under the assumptions in the
introduction i.e. when the random variables admit a density, one has
$\P(\omega_0\in\{-r,r\})=0$.\\
We will use the following classical two-sided bound on the integrated
density of states obtained using Dirichlet-Neumann bracketing (see
e.g.~\cite{MR94h:47068,MR1935594}): there exists $C>0$ such that, for
$L\geq 1$, one has
\begin{equation}
  \label{eq:76}
  \frac1L\P\{E_{D,L}(\omega)\leq E\}\leq N(E)\leq
  C\,\P\{E_{N,L}(\omega)\leq E\}
\end{equation}
where
\begin{itemize}
\item $E_{D,L}(\omega)$ is the ground state of $H_{\omega,L}^D$ with
  Dirichlet boundary conditions,
\item $E_{N,L}(\omega)$ is the ground state of $H_{\omega,L}^D$ with
  Neumann boundary conditions,
\item $C$ is a constant depending only on $\|V\|_\infty$.
\end{itemize}
We now use it to obtain the initial length scale estimate needed in
addition to the Wegner estimate to apply~\cite[Theorem
5.4]{MR2509111}. Indeed, by~\eqref{eq:62} and~\eqref{eq:76}, for any
$a>0$ and $b \in (0,1)$ there exists $\tilde L=\tilde L(a,b)$ such
that, for all $L\geq\tilde L$.
\begin{equation*}
  \P(H^D_{\omega,L} \:\mbox{(with Dirichlet b.c.) has an eigenvalue less than}\:
  \inf\Sigma^D+L^{-b}) \leq L^{-a}.
\end{equation*}
Using standard Combes-Thomas estimates (see e.g.~\cite{MR1935594}),
this implies that there exists $C>0$ such that, with probability, as
least $1-L^{-a}$, one has
\begin{equation*}
  \sup_{E\leq \inf\Sigma^D+L^{-b}/2}\|\chi_x
  (H^D_{\omega,L}-E)^{-1}\chi_y\|\leq e^{-L^{-b}|x-y|/C}
\end{equation*}
where $\chi_x=\car_{[x-1/2,x+1/2]}$.\\
This estimate immediately yields that assumption (5.7)
of~\cite[Theorem 5.4]{MR2509111}) is satisfied in some neighborhood of
$\inf\Sigma^D$ for the model $H_\omega^D$ considered in the
introduction. Thus, we obtain Theorem~\ref{thr:2}.\vskip.1cm
\noindent Let us now return to Theorems~\ref{thr:12}
and~\ref{thr:8}. Before proving these results, let us give a more
precise result under a simple additional assumption on the random
variables $(\omega_n)_{n\in\Z}$. We prove
\begin{Th}
  \label{thr:11}
  Assume that the common distribution of the displacements
  $(\omega_n)_{n\in\Z}$ satisfies $\P(\omega_0=-r)+\P(\omega_0=r)=0$
  and
  \begin{equation}
    \label{eq:74}
    \lim_{\varepsilon\to0^+}
    \frac{\log\left|\log\P(\omega_0\in[-r,-r+\varepsilon])+
        \log\P(\omega_0\in[r-\varepsilon,r])\right|}
    {\log|\log\varepsilon|}=1.
  \end{equation}
  Then, one has
  \begin{equation}
    \label{eq:75}
    \lim_{\substack{E\to\inf\Sigma^D\\E>\inf\Sigma^D}}
    \frac{\log|\log N(E)|}{\log|\log(E-\inf\Sigma^D)|}=2.
  \end{equation}
\end{Th}
\noindent Up to terms of smaller order, the limit~\eqref{eq:75} should
be interpreted as
\begin{equation*}
  N(E)\sim e^{-C|\log(E-\inf\Sigma^D)|^2},
\end{equation*}
and assumption~\eqref{eq:74} as, for some $n_+>n_->0$ and
$\varepsilon$ positive sufficiently small, one has
\begin{equation}
  \label{eq:83}
  \varepsilon^{n_-}
  \leq\P(\omega_0\in[-r,-r+\varepsilon])\leq\varepsilon^{n_+}\quad
  \text{ and }\quad    \varepsilon^{n_-}
  \leq \P(\omega_0\in[r-\varepsilon,r])
  \leq\varepsilon^{n_+}.
\end{equation}
When the common distribution of the $(\omega_n)_{n\in\Z}$ is even, a
lower bound for $N(E)$ was obtained in~\cite[section 4]{MR2503172}
(even though it was not stated explicitly); it was of size
$e^{-C|\log(E-\inf\Sigma^D)|^3}$.
\begin{Rem}
  \label{rem:8} In Theorems~\ref{thr:8} and~\ref{thr:11}, the
  smoothness assumption on $V$ can be relaxed quite a bit (see
  e.g.~\cite{MR2503172}).
\end{Rem}
\noindent Let us now turn to the proof of
Theorems~\ref{thr:12},~\ref{thr:8} and~\ref{thr:11}.  For $L>0$,
consider $H^D_{\omega,L}$ the operator $H^D_\omega$ restricted to the
interval $[-L+1/2,L+1/2]$; the boundary conditions will be made
precise below.\\
Our main tools to prove Theorems~\ref{thr:12},~\ref{thr:8}
and~\ref{thr:11} are the two following lemmas
\begin{Le}
  \label{le:4}
  There exists $C>1>c>0$, $\tau\in(0,1)$ and $\varepsilon_0>0$ such
  that, for $\varepsilon\in(0,\varepsilon_0)$ and
  $L\geq\varepsilon_0^{-1}$, one has
  \begin{multline}
    \label{eq:85}
    \P\{E_{D,L}(\omega)\leq\inf\Sigma^D+C(\varepsilon+\tau^{2L})\}
    \\\geq \left[\P\left(\omega_0\in\left[-r,-r+
          c\varepsilon\right]\right)\,
      \P\left(\omega_0\in[r-c\varepsilon,r]\right)\right]^L.
  \end{multline}
\end{Le}
\noindent and
\begin{Le}
  \label{le:5}
  Set $p:=\P(\omega_0\in[-r,0]\in(0,1)$. Then, there exists $C>1$
  such, for $\varepsilon\in(0,1)$ and $L\geq1$, one has

  \begin{equation}
    \label{eq:86}
    \P\{E_{N,L}(\omega)\leq\inf\Sigma^D+\varepsilon\}
    \leq\sum_{k=0}^L
    \sum_{\substack{K\subset\{-L+1,\cdots,L\}\\ \#K=k}}
    P_{\mathcal{K},L}(\varepsilon)
  \end{equation}
  where
  \begin{multline}
    \label{eq:87}
    P_{\mathcal{K},L}(\varepsilon):=\sum_{m=0}^L
    \prod_{n\in\mathcal{K}} \P\left(\omega_0\in\left[-r,-r+CL\,
        e^{C|n-m|}\varepsilon\right]\right) \\
    \prod_{n\not\in\mathcal{K}}
    \P(\omega_0\in[r-CL\,e^{C|n-m|}\varepsilon,r]).
  \end{multline}
\end{Le}
\noindent Let us now show how these lemmas are used to prove
Theorems~\ref{thr:12},~\ref{thr:8} and~\ref{thr:11}.\\
We start with Theorem~\ref{thr:12} and the lower bound in
Theorem~\ref{thr:11}. Pick $\varepsilon$ positive small and $L$ such
that
\begin{equation}
  \label{eq:80}
  L-1\leq\alpha|\log\varepsilon|\leq L.
\end{equation}
where $\alpha>0$. If we pick $\alpha\geq(2\log\tau)^{-1}$ then
$\tau^{2L}\leq\varepsilon$. Under the assumptions of
Theorem~\ref{thr:12}, the bound~\eqref{eq:86} and the lower bound
in~\eqref{eq:80} yield, for some $\nu\in(0,1)$,
\begin{equation*}
  N(\inf\Sigma^D+2C\varepsilon)\geq
  \nu^{|\log\varepsilon|}=\varepsilon^{|\log\nu|}.
\end{equation*}
One completes the proof of Theorem~\ref{thr:12} by taking
$n>|\log\nu|$.\\
The lower bound in~\eqref{eq:75} in Theorem~\ref{thr:8} is obtained in
the same way. For $\alpha$ in~\eqref{eq:80} sufficiently large, we
obtain that
\begin{multline*}
  \log|\log N(\inf\Sigma^D+2C\varepsilon)|\\\geq \log L
  +\log\left|\log \P\left(\omega_0\in\left[-r,-r+
        c\varepsilon\right]\right)+\log
    \P\left(\omega_0\in[r-c\varepsilon,r]\right) \right|.
\end{multline*}
Thus, assumption~\eqref{eq:74} and the bound~\eqref{eq:80} immediately
yield the lower bound in~\eqref{eq:75}.\\
Let us now turn to the proof of Theorem~\ref{thr:8} and the upper
bound in Theorem~\ref{thr:11}. We again pick $\varepsilon$ positive
small and $L$ such that~\eqref{eq:80} be satisfied for some
$\alpha>0$. Now $\alpha$ is chosen so small that $C \alpha<1/4$ where
$C$ is given by Lemma~\ref{le:5}. Thus, for $\varepsilon$ small and
$(n,m)\in\{-L+1,\cdots,L\}^2$, one has $CL\,e^{C|n-m|}\varepsilon\leq
\sqrt{\varepsilon}$ and~\eqref{eq:86} becomes
\begin{multline}
  \label{eq:91}
  \P\{E_{N,L}(\omega)\leq\inf\Sigma^D+\varepsilon\} \\\leq L^2\, 2^L\,
  \left(\P\left(\omega_0\in\left[-r,-r+
        \sqrt{\varepsilon}\right]\right)+
    \P\left(\omega_0\in[r-\sqrt{\varepsilon},r]\right) \right)^{2L}
\end{multline}
Under the assumptions of Theorem~\ref{thr:12} or Theorem~\ref{thr:11},
for $\varepsilon$ small, we get
\begin{multline*}
  \log\P\{E_{N,L}(\omega)\leq\inf\Sigma^D+\varepsilon\}
  \\\leq-\alpha|\log\varepsilon|\log\left(\P\left(\omega_0\in\left[-r,-r+
        \sqrt{\varepsilon}\right]\right)+
    \P\left(\omega_0\in[r-\sqrt{\varepsilon},r]\right) \right).
\end{multline*}
This immediately gives~\eqref{eq:62} under the assumptions of
Theorem~\ref{thr:12} and the upper bound in~\eqref{eq:75} under those
of Theorem~\ref{thr:11}. Hence, the proofs of Theorem~\ref{thr:12} and
Theorem~\ref{thr:11} are complete.
\begin{proof}[The proof of Lemma~\ref{le:4}]
  As $V$ is smooth and compactly supported, we know that there exists
  $c\in(0,1)$, such that for any admissible $(\omega_n)_{-L+1\leq
    n\leq L}$ and $(\omega'_n)_{-L+1\leq n\leq L}$, one has
  \begin{equation}
    \label{eq:78}
    \begin{split}
      \|H_{\omega,L}^D-H_{\omega',L}^D\|&=\sup_{[-L+1/2,L+1/2]}
      \left|\sum_{n\in\Z} V(\cdot-n-\omega_n)-\sum_{n\in\Z}
        V(\cdot-n-\omega'_n)\right|\\&\leq c^{-1}\sup_{-L+1\leq n\leq
        L}|\omega_n-\omega'_n|.
    \end{split}
  \end{equation}
  Here, $\|\cdot\|$ denotes the operator norm and the estimate does
  not depend on the boundary conditions used to define
  $H_{\omega',L}^D$ (provided we use the same boundary conditions for
  $H_{\omega',L}^D$ and $H_{\omega,L}^D$).\\
  Recall that, for $a\in [-r,r]$, we have defined $H_1(a) = -\Delta +
  q(x-a)$ on $L^2(-1/2,1/2)$ with Neumann boundary condition and
  $E_0(a) = \inf \sigma(H_1(a))$ to be the lowest eigenvalue of
  $H_1(a)$. Let $\psi_0(a;x)$ be the associated positive ground state.
  Note that, by symmetry, one has
  $\psi_0(-a;x)=\psi_0(a;-x)$. By~\cite[Lemma 3.2]{MR2503172}, we know
  that $\psi_0(a;-1/2)\not=\psi_0(a;1/2)$ as $a\mapsto E_0(a)$ is
  supposed not to be constant. For $a=r$, assume that
  \begin{equation*}
    0<\frac{\psi_0(r;-1/2)}{\psi_0(r;1/2)}:=\tau<1
  \end{equation*}
  If this is not the case, in the construction that follows, we
  invert the parts of $r$ and $-r$.\\
  By the results of~\cite{MR2503172}, we know that
  $E(-r)=E(r)=\inf\Sigma^D$. \\
  Consider the event
  \begin{equation*}
    \Omega_{L,\varepsilon}=\left\{
      \begin{aligned}
        &\forall n\in\{-L+1,0\},\quad
        &|\omega_n+r|\leq\varepsilon\\
        &\forall n\in\{1,L\},\quad &|\omega_n- r|\leq\varepsilon
      \end{aligned}
    \right\};
  \end{equation*}
  The $(\omega_n)_{n\in\Z}$ being independent, the probability of this
  event is bounded from below
  \begin{equation}
    \label{eq:77}
    \P(\Omega_{L,\varepsilon})\geq
    \left[\P(\omega_0\in[-r,-r+\varepsilon])
      \P(\omega_0\in[r-\varepsilon,r])\right]^L.
  \end{equation}
  For the realization $(\omega^r_n)_{-L+1\leq n\leq L}$ defined by
  $\omega^r_n=-r$ if $n\in\{-L+1,0\}$ and $\omega^r_n=r$ if
  $n\in\{1,L\}$, we know (see~\cite{MR2503172}) that
  $\psi_{\omega^r,L}$, the normalized positive ground state of
  $H_{\omega,L}^D$ with Neumann boundary conditions, is given by
  \begin{equation*}
    \psi_{\omega^r,L}(x)=\frac1{C_0}
    \begin{cases}
      \tau^{-n}\,\psi_0(r;n-x) \text{ if }n\in\{-L+1,0\}\\
      \tau^{n-1}\,\psi_0(r;x-n)\text{ if }n\in\{1,L\}
    \end{cases}
    \quad\text{for }-\frac12\leq x-n\leq\frac12
  \end{equation*}
  where
  \begin{equation*}
    C^2_0=\sum_{n=0}^{L-1}\tau^{2n}\int_{-1/2}^{1/2}|\psi_0(r;x)|^2dx
    =\frac{1-\tau^{2L}}{1-\tau^2}>1.
  \end{equation*}
  Here, we have used the symmetries of $(a,x)\mapsto\psi_0(a;x)$ and
  the fact that it is normalized.\\
  Pick $\chi:\ (-L+1/2,L+1/2)\to\R^+$ smooth such that
  $0\leq\chi\leq1$, $\chi\equiv1$ on $(-L+1,L-1)$ and it vanishes
  identically near $L+1/2$ and $-L+1/2$.  Consider the function
  $\phi=\chi\psi_{\omega^r,L}$. It satisfies Dirichlet boundary
  conditions at $L+1/2$ and $-L+1/2$.  Moreover, using~\eqref{eq:78},
  for $\omega\in\Omega_{r,c\varepsilon}$ (recall that $c$ is defined
  in~\eqref{eq:78}), one computes that
  $1-\tau^{2L}C_0^{-2}\leq\|\phi\|^2\leq 1$ and there exists $C>0$
  such that
  \begin{equation}
    \label{eq:79}
    \|(H_{\omega,L}^D-E_0(r))\phi\|^2\leq C\left(\tau^{2L}
      +\varepsilon\right)^2\leq C^2\left(\tau^{2L}
      +\varepsilon\right)^2\|\phi\|^2.
  \end{equation}
  This and estimate~\eqref{eq:77} immediately yields~\eqref{eq:85} and
  completes the proof of Lem\-ma~\ref{le:4}.
\end{proof}
\begin{proof}[The proof of Lemma~\ref{le:5}]
  We are going to rely on the analysis done for the Lifshitz tails
  regime in~\cite[section 3]{1007.2483}. Define the random variable
  $\omega^e_0$ and $\tilde\omega_0$ as
  $\omega^e_0=r\,(\car_{\omega_0>0}-\car_{\omega_0\leq0})$ and
  $\tilde\omega_0=|\omega_0-\omega_0^e|$ conditioned on
  $\omega^e_0$. Note that, under the assumptions of
  Theorems~\ref{thr:8} and~\ref{thr:11}, $\tilde\omega_0$ is not
  identically vanishing. In the same way, for any $n\in\Z$, define
  $\omega^e_n$ and $\tilde\omega_n$. Then, though not stated directly
  in this way, the following result is proved in~\cite[section
  3]{1007.2483}
  \begin{Le}[\cite{1007.2483}]
    \label{le:6}
    There exists $C>0$ such that
    \begin{equation*}
      H_{\omega,L}^D-\inf\Sigma^D\geq \frac1C\left(
        H_{\omega^e,L}^D-\inf\Sigma^D+V_{\tilde\omega,L}\right)
    \end{equation*}
    where
    \begin{equation*}
      V_{\tilde\omega,L}(x)=\sum_{n=-L+1}^L
      \tilde\omega_n\car_{[-1/2,1/2]}(x-n).
    \end{equation*}
  \end{Le}
  \noindent Using this decomposition and recalling that
  $p=\P(\omega_0\in[-r,0]\in(0,1)$, we can write
  \begin{equation}
    \label{eq:89}
    \P\{E_{N,L}(\omega)\leq\inf\Sigma^D+\varepsilon\}
    \leq\sum_{k=0}^L
    p^k(1-p)^{L-k}\sum_{\substack{K\subset\{-L+1,\cdots,L\}\\ \#K=k}}
    \tilde P_{\mathcal{K},L}(\varepsilon)
  \end{equation}
  where
  \begin{equation*}
    \tilde P_{\mathcal{K},L}(\varepsilon)=\P
    \left\{
      \begin{aligned}
        \exists \varphi\in C^1,\ \|\varphi\|=1\text{and }\exists
        E\in[0,C\varepsilon]\text{ s.t. }\\
        \left(H_{\omega^e,L}^D-\inf\Sigma^D+V_{\tilde\omega,L}-E\right)\varphi=0
      \end{aligned}
      \left|
        \begin{aligned}
          \omega^e_n&=-r\text{ for }n\in\mathcal{K}
          \\\omega^e_n&=r\text{ for }n\not\in\mathcal{K}
        \end{aligned}
      \right.
    \right\}.
  \end{equation*}
  Lemma~\ref{le:13} guarantees that there exists $C>0$ (independent of
  $L$ and the realization $\omega$) such that, if $\varphi$ is a
  solution to
  $(H_{\omega^e,L}^D-\inf\Sigma^D+V_{\tilde\omega,L}-E)\varphi=0$,
  \begin{equation*}
    \forall(m,n)\in\{-L+1,\cdots,L\},\quad
    \int_{-1/2}^{1/2}|\varphi(x-n)|^2dx\leq
    e^{C|m-n|}\int_{-1/2}^{1/2}|\varphi(x-m)|^2dx.
  \end{equation*}
  If $\varphi$ is normalized, we know that one has
  $\D\int_{-1/2}^{1/2}|\varphi(x-n)|^2dx\geq (2L)^{-1}$ for some
  $n\in\{-L+1,\cdots,L\}$.\\
  As $H_{\omega^e,L}^D-\inf\Sigma^D\geq0$, these two properties imply
  that
  \begin{equation*}
    \begin{split}
      \tilde P_{\mathcal{K},L}(\varepsilon)&\leq \sum_{m=-L+1}^L\P
      \left\{ \sum_{n=-L+1}^L\tilde\omega_ne^{-C|m-n|}\leq 2CLE \left|
          \begin{aligned}
            \omega^e_n&=-r\text{ for }n\in\mathcal{K}
            \\\omega^e_n&=r\text{ for }n\not\in\mathcal{K}
          \end{aligned}
        \right.  \right\}\\&\leq \sum_{m=-L+1}^L
      \prod_{n\in\mathcal{K}} \P\left(\omega_0\in\left[-r,-r+2CL\,
          e^{C|n-m|}\varepsilon\right]|\omega^e_n=-r\right)
      \\&\hskip5cm \prod_{n\not\in\mathcal{K}}
      \P(\omega_0\in[r-2CL\,e^{C|n-m|}\varepsilon,r]|\omega^e_n=r).
    \end{split}
  \end{equation*}
  Using the definition of $(\omega^e_n)_{n\in\Z}$, we immediately
  obtain the bound~\eqref{eq:86} and thus complete the proof of
  Lemma~\ref{le:5}.
\end{proof}
\section{Appendix}
\label{sec:appendix}
\noindent In this appendix, we collect various technical results that
were used in our study.
\subsection{Some results on differential equations}
\label{sec:some-results-diff}
We recall some standard estimates on ordinary differential equations
that are immediate consequences of equations~\eqref{eq:27}
and~\eqref{eq:28}, and, presumably well known (see
e.g.~\cite{MR89b:47070,0912.3568}). We use the notation of
section~\ref{sec:inverse-tunneling}.
\begin{Le}
  \label{le:13}
  There exists a constant $C>0$ (depending only on $\|q\|_\infty$)
  such that, for $u$ a solution to $Hu=0$ (see~\eqref{eq:25}), if
  $I(x):=[x-1/2,x+1/2]\cap[0,\ell]$, one has
  \begin{gather}
    \label{eq:23}
    \forall x\in[0,\ell],\quad \frac1C\int_{I(x)}u^2(y)dy\leq
    r^2_u(x)\leq C\int_{I(x)}u^2(y)dy,\\\label{eq:30} \forall
    x\in[0,\ell],\quad \min_{I(x)} r_u\leq\max_{I(x)} r_u\leq
    C\,\min_{I(x)} r_u, \\\label{eq:26} \forall x\in[0,\ell],\quad
    \|\sin(\varphi_u(\cdot))\|_{L^2(I(x))}\geq\frac1C.
  \end{gather}
\end{Le}
\begin{Le}
  \label{le:15}
  Let $\delta\varphi$ be a solution to the
  equation~\eqref{eq:31}. There exists $C>0$ (depending only on
  $\|q\|_\infty$) such that, for $x_0\in[0,\ell]$, one has
  \begin{equation*}
    \forall x\in[0,\ell],\quad    |\sin(\delta\varphi(x))|\leq
    [|\sin(\delta\varphi(x_0))|+E\ell]e^{C|x-x_0|}.
  \end{equation*}
\end{Le}
\begin{proof}
  Write $\D s(x)=|\sin(\delta\varphi(t))|$ and note that, integrating
  equation~\eqref{eq:31} implies that
  \begin{equation*}
    s(x)\leq s(x_0)+E\ell+C\int_{x_0}^xs(t)dt.
  \end{equation*}
  The statement of Lemma~\ref{le:15} then follows from Gronwall's
  Lemma (see e.g.~\cite{MR923320}).
\end{proof}
\begin{Le}
  \label{le:17}
  There exists $\eta_0>0$ depending only on $\|q\|_\infty$ such that,
  for $\eta\in(0,\eta_0)$ and $\varphi_u$, a solution to
  equation~\eqref{eq:27}, one has
  \begin{enumerate}
  \item if $y<y'$ are such that $\D\max_{x\in
      [y,y']}|\sin\varphi_u(x)|\leq\eta$, then $|y-y'|\leq
    \eta/\eta_0$;
  \item if $|\sin(\varphi_u(y))|\leq\eta$ then, for
    $4\eta\leq|x-y|\leq\sqrt{\eta}$, one has
    \begin{equation*}
      |\sin(\varphi_u(y))|\geq|x-y|/2.
    \end{equation*}
  \item if $y<y'$ are such that
    \begin{equation*}
      |\sin\varphi_u(y)|=
      |\sin\varphi_u(y')|=\eta\text{ and }\min_{x\in
        [y,y']}|\sin\varphi_u(x)|\geq\eta
    \end{equation*}
    then $|y-y'|\geq (\eta_0-\eta)\eta_0$.
  \end{enumerate}
\end{Le}
\begin{proof}
  First, by equation~\eqref{eq:27}, for some $C>0$ depending only on
  $\|q\|_\infty$, one has $|\varphi'_u(x)|\leq C$ and, if
  $|\sin(\varphi_u(x))|\leq\eta$ then $1-C\eta^2\leq
  |\cos\varphi_u(x)|\varphi'_u(x)$. Pick $\eta_0\in(0,1)$ such that
  $1-C\eta_0^2\geq1/2$.\\
  To prove point (1), consider $y<y'$ such that $\D\max_{x\in
    [y,y']}|\sin\varphi_u(x)|\leq\eta$. As $\eta<\eta_0<1$,
  $\cos\varphi_u(x)$ does not change sign on $[y,y']$. Thus, one
  computes
  \begin{equation*}
    2\eta\geq|\sin\varphi_u(y')-\sin\varphi_u(y)|=\int_y^{y'}
    |\cos\varphi_u(x)|\varphi'_u(x)dx\geq |y-y'|/2. 
  \end{equation*}
  This proves (1) possibly diminishing the value of $\eta_0$.\\
  To prove point (2), as by equation~\eqref{eq:27}, for some $C>0$
  depending only on $\|q\|_\infty$, one has $|\varphi'_u(x)|\leq C$,
  there exists $\eta_0>0$ such that, for $\eta\in(0,\eta_0]$, if
  $|\sin(\varphi_u(y))|\leq\eta$, one has
  $|\sin(\varphi_u(x))|\leq\eta_0$ for $|x-y|\leq\eta_0$. Thus, at the
  possible cost of reducing $\eta_0$, $x\mapsto|\cos(\varphi_u(x))|$
  stays larger than $9/10$ on $[y-\eta_0,y+\eta_0]$, and, by
  equation~\eqref{eq:27}, one has $d/dx[\sin(\varphi_u(x))]\geq 3/4$
  on $[y-\eta_0,y+\eta_0]$. This, the assumption
  $|\sin(\varphi_u(y))|\leq\eta$ and the Taylor formula immediately
  entail point (2).\\
  To prove point (3), note that, as $|\varphi'_u(x)|\leq C$, for
  $y<z<y+(\eta_0-\eta)/C$, one has
  $|\sin(\varphi_u(z))|\leq\eta_0$. Thus, $x\mapsto\cos\varphi_u(x)$
  keeps a constant sign on the interval
  $[y,\min(y',y+(\eta_0-\eta)/C)]$. Moreover, as $\D\min_{x\in
    [y,y']}|\sin\varphi_u(x)|\geq\eta$, so does
  $x\mapsto\sin\varphi_u(x)$ and both signs are the same. Thus, for
  $y<z<y+(\eta_0-\eta)/C$, we know that
  \begin{equation*}
    \begin{split}
      |\sin\varphi_u(z)|&=|\sin\varphi_u(y)|+\int_y^z
      |\cos\varphi_u(x)|\varphi'_u(x)dx\\&
      \geq\eta+\sqrt{1-(\eta')^2}(z-y)/2>\eta.
    \end{split}
  \end{equation*}
  Hence, one has $y'>y+(\eta_0-\eta)/C$. This proves (2) at the
  expense of possibly changing $\eta_0$ again. This completes the
  proof of Lemma~\ref{le:17}.
\end{proof}

\def\cprime{$'$} \def\cydot{\leavevmode\raise.4ex\hbox{.}} \def\cprime{$'$}

\end{document}